\documentclass[draft, jgrga]{AGUTeX}

  \usepackage[]{graphicx}
  \setkeys{Gin}{draft=false}
\authorrunninghead{Phipps, P.H.}

\titlerunninghead{\textit{Juno} Radio occultation experiment}

\begin{document}

%% ------------------------------------------------------------------------ %%
%
%  TITLE
%
%% ------------------------------------------------------------------------ %%

\title{Radio occultations of the Io plasma torus by \textit{Juno} are feasible}

% e.g., \title{Terrestrial ring current:
% Origin, formation, and decay $\alpha\beta\Gamma\Delta$}
%

%% ------------------------------------------------------------------------ %%
%
%  AUTHORS AND AFFILIATIONS
%
%% ------------------------------------------------------------------------ %%

%Use \author{\altaffilmark{}} and \altaffiltext{}

% \altaffilmark will produce footnote;
% matching \altaffiltext will appear at bottom of page.

 \authors{Phillip H. Phipps,\altaffilmark{1}
 Paul Withers,\altaffilmark{1,2}}%

\altaffiltext{1}{Department of Astronomy,
Boston University, Boston, Massachusetts, USA.}

\altaffiltext{2}{Center for Space Physics, Boston University, Boston, Massachusetts, USA.}

%\altaffiltext{3}{Department of Space Sciences, University of
%Michigan, Ann Arbor, Michigan, USA.}

%\altaffiltext{4}{Division of Hydrologic Sciences, Desert Research
%Institute, Reno, Nevada, USA.}

%\altaffiltext{5}{Dipartimento di Idraulica, Trasporti ed
%Infrastrutture Civili, Politecnico di Torino, Turin, Italy.}

%% ------------------------------------------------------------------------ %%
%
%  ABSTRACT
%
%% ------------------------------------------------------------------------ %%

% >> Do NOT include any \begin...\end commands within
% >> the body of the abstract.

\begin{abstract}
%PENDING
%XXX1 - JGR has 250 word limit, previous version was 350 words or so.
%http://publications.agu.org/author-resource-center/text-requirements/POUNDSIGNabstract
%
%What is the main overarching issue that the study described in the paper will help resolve in the future?
%Which critical hurdle towards the main overarching issue has the study described in the paper cleared?
%What is new in the study that enabled it to clear the hurdle?
%What new is learned in the study?
%What is the implication of the new results in the context of future endeavors towards the main overarching issue? 
%
%From Masaki Fujimoto, JGR Space Physics editor, in rejecting at least two of my manuscripts without sending them out to review.
%
 The flow of material from Io's volcanoes into the Io plasma torus, out into the magnetosphere, and along field lines into Jupiter's upper atmosphere is not adequately understood.
The lack of observations of spatial and temporal variations in the Io plasma torus impedes attempts to understand the system as a whole.
Here we propose that radio occultations of the Io plasma torus by the \textit{Juno} spacecraft can measure plasma densities in the Io plasma torus.
We find that the line-of-sight column density of plasma in each of the three regions of the Io plasma torus (cold torus, ribbon, and warm torus) can be measured with uncertainties of 10\%.
We also find that scale heights describing the spatial variation in plasma density in each of these three regions can be measured with similar uncertainties.
Such observations will be sufficiently accurate to support system-scale studies of the flow of plasma through the magnetosphere of Jupiter. 
\end{abstract}

%% ------------------------------------------------------------------------ %%
%
%  BEGIN ARTICLE
%
%% ------------------------------------------------------------------------ %%

% The body of the article must start with a \begin{article} command
%
% \end{article} must follow the references section, before the figures
%  and tables.

\begin{article}

%% ------------------------------------------------------------------------ %%
%
%  TEXT
%
%% ------------------------------------------------------------------------ %%

\section{\label{sec:intro}Introduction}
Volcanic eruptions on the innermost Galilean satellite, Io, are the main source of plasma in Jupiter's magnetosphere. Io orbits Jupiter in the plane of the planet's rotational equator at a distance of 5.9 $R_J$. Volcanic activity on Io delivers neutral gas into the Jupiter system at a rate of about 1 tonne per second \citep[e.g.][]{2011Bagenal}. Plasma is produced from these neutrals via electron collisions on timescales of 2--5 hours \citep{1988Smyth,1992Smyth,2004Thomas}. Plasma is also transferred into the magnetosphere as Jupiter's rapidly rotating magnetic field picks up ions from Io's ionosphere. The relative importance of these two processes is not currently known \citep{2004Thomas}.

%Another source of plasma is the ionosphere of Io produced by photo-ionization but it is unknown if ionization at Io or in the torus away from Io is the dominant source of plasma \citep{2004Thomas}.}% THIS SENTENCE NEEDS FIXING ONCE I KNOW WHAT IT IS TRYING TO SAY, ALSO NEEDS TO BE SUPPORTED BY REFERENCES.} %PUT MORE REFERENCES HERE
%These neutral atoms, predominantly oxygen and sulfur, are ionized on timescales of a few hours by electron collisions and charge exchange with ions. Specifically, the timescales for ionization by electron collisions are approximately 13.6 hours and 1.6 hours for oxygen and sulfur atoms, respectively. The timescales for ionization by charge exchange are approximately 3.5 hours and 2.5 hours for oxygen and sulfur atoms, respectively. These timescales were calculated using rate coefficients from \citet{2003Delamere} and \citet{2005Steffl} with neutral oxygen and sulfur densities of 29 and 6 cm$^{-3}$, respectively.
%PLEASE SHOW ME AT LEAST ONE OF THESE CALCULATIONS AND SOURCES SO I CAN BE REASSURED IT MAKES SENSE., called the Io plasma torus, 
Once ionized, these particles are affected by electromagnetic forces in addition to gravitational and centrifugal forces. These forces disperse the Io-genic plasma away from Io, but do not do so uniformly in all directions. Instead, the plasma is initially confined to a torus that is centered on the centrifugal equator at Io's orbital distance (5.9 $R_J$), called the Io plasma torus (IPT).
The centrifugal equator is the locus of points on a given field line which are located at the greatest distance from the rotation axis
\citep{1974Hill,2002Dessler,2004Khurana}.
The torus is centered in this plane since this plane is where an ion trapped on a field line has the minimum centrifugal potential.
The axis of the centrifugal equator lies between Jupiter's rotational and magnetic axes and is therefore tilted towards the magnetic axis at 200$^{\circ}$ longitude (System III).
The angle between the rotational and centrifugal axes is 2/3 the angle between the rotational and magnetic axes \citep{1974Hill}. 
Since Jupiter's magnetic axis is 9.6 degrees from the rotational axis, the centrifugal axis that defines the plane of the IPT is tilted by 6.4 degrees from the rotational axis and 3.2 degrees from the magnetic axis \citep{2004Thomas}. 
The jovian magnetic field is not perfectly dipolar. Consequently, representing the centrifugal equator as a plane is an approximation.
However, doing so is sufficient for many purposes.
%NEED TO SHOW ME EXACTLY WHERE THESE NUMBERS COME FROM, EXPLAIN ``ADAPTED'', AND EXPRESS WITH APPROPRIATE NUMBER OF SIGNIFICANT FIGURES. %I DON'T LIKE THE SOUND OF THIS DEFINITION, I THINK IT OUGHT TO BE EXPRESSED DIFFERENTLY, LET'S DISCUSS in which the magnetic and centrifugal forces sum to zero 
%CONSIDER A BUCKET WITH TWO HOLES, ONE BIG AND ONE SMALL. THE BIG HOLE DRAINS THE BUCKET ON A TIMESCALE OF A FEW HOURS. THE SMALL HOLE DRAINS THE BUCKET ON A TIMESCALE OF 20-80 DAYS. AFTER A FEW HOURS, NO WATER REMAINS TO DRIP OUT OF THE SMALL HOLE. I HAVE FORGOTTEN THE EXPLANATION YOU GAVE ME EARLIER FOR WHY THIS ANALOGY FAILS. AS YOU CAN SEE, THE TEXT DOESN'T MAKE SENSE TO ME.
%Approximately two-thirds of the plasma I CHANGED MATERIAL TO PLASMA, WAS THAT SIGNIFICANT? in the Io plasma torus is lost via charge exchange on the order of a few hours. 
%The plasma drifts radially outward via flux tube interchange on a timescale of around 20--80 days \citep{1981Hill,2011Bagenal}, and is distributed throughout Jupiter's middle and outer magnetosphere.

%The cross-section of the torus is approximately circular with a radius of 1--2 $R_{J}$ \citep{1981Bagenal}.
%The torus does not have a circular cross-section. Instead, it has an extent of around 4 $R_J$ normal to the equatorial plane and 3.1 $R_J$ in the equatorial plane.
Plasma is lost from the IPT via flux tube interchange on timescales of around 20-80 days \citep{1981Hill,2011Bagenal}. Flux tube interchange causes  plasma to drift radially outward, which distributes plasma throughout Jupiter's middle and outer magnetosphere.  
%Most plasma in Jupiter's magnetosphere passes through the Io plasma torus at an early stage in its life cycle.
The dispersal of plasma from the IPT into the rest of the magnetosphere is the main process that provides plasma to the rest of Jupiter's magnetosphere.
Therefore spatial and temporal variations in the IPT can  ultimately affect the distribution and dynamics of plasma throughout Jupiter's entire magnetosphere \citep{2012Bonfond,2014Payan}. 

%\textbf{In fact, \citet{2004Nozawa} showed a corellation of sulfur ion densities between the inner and outer magnetosphere of Jupiter. Thus, it is important to observe temporal and spatial variations in the torus.} 
%Such variations ultimately affect the distribution and dynamics of plasma throughout Jupiter's entire magnetosphere \citep{2012Bonfond,2014Payan}. 

%Several key questions concerning the Io plasma torus address the production, maintenance, and loss of plasma. What is the variability for production of plasma at Io?, How do changes in Io's volcanic activity affect the torus? and What is the temporal and spatial distribution of the flux-tube interchange loss mechanism? 
%THIS MAKES LITTLE SENSE. IF 67 PROTONS ARE LOST IN 200 SECONDS, THEN THE REMAINING 33 ARE NOT GOING TO HANG AROUND FOR 20 DAYS. LET'S DISCUSS THE SOURCE OF THIS STATEMENT.

%THERE'S AN OPPORTUNITY HERE TO SKETCH OUT THE SCIENCE SCENARIO WE DISCUSSED OF GROUND-BASED IR VOLCANIC ACTIVITY OBSERVATIONS, GROUND-BASED SODIUM CLOUD OBSERVATIONS, JUNO IPT RADIO OCCULTATIONS, ALL THE OTHER JUNO OBSERVATIONS PROVIDING AN INTEGRATED, COMPREHENSIVE VIEW OF THE FLOW OF STUFF FROM IO'S VOLCANOES TO JUPITER AURORAL EMISSIONS. LET'S GET THE REST OF THE TEXT SORTED OUT BEFORE THINKING ABOUT THAT. MIGHT FIT BETTER IN CONCLUSIONS SECTION.
%Answers to these questions are important for the understanding of interactions with Io and Jupiter's magnetosphere.

There are several different ways in which remote sensing and in situ observations can measure conditions in the IPT. This article focuses on remote sensing observations of the IPT by radio occultations. These observations can monitor temporal and spatial variations in the density and temperature of the IPT.

%can be studied remotely by spacecraft via radio occultations.
A radio occultation occurs when an object, here the IPT, comes between the transmitter and receiver of a radio signal.
Properties of the radio signal are affected by the radio signal's propagation through the plasma in the torus.
%3. RADIO OCCULTATIONS OF THE PLASMA TORUS, WHAT THEY ARE, WHAT THEY MEASURE, WHAT PROPERTIES CAN BE INFERRED FROM SUCH MEASUREMENTS.
%A spacecraft can use radio occultation to remotely measure plasma densities in the Io torus.
%A radio occultation can occur when plasma comes between a receiver and the spacecraft. 
Refraction of the radio signal as it passes through the plasma of the IPT causes a change in the frequency of the received signal due to the Doppler effect.
The line-of-sight integrated plasma density, also known as the total electron content (TEC), of the IPT can be determined from the measured shift in the received frequency \citep[e.g.][]{2014Withers}.
%This is also known as the total electron content.
The most suitable scenario for a radio occultation observation of the IPT involves a spacecraft in a polar orbit around Jupiter with periapsis within Io's orbit.
A polar orbit ensures that the line of sight between the spacecraft and Earth is approximately parallel to the torus equator and sweeps through the entire cross-section of the IPT.
A periapsis within Io's orbit ensures that the line of sight between the spacecraft and Earth passes through the torus once, not twice, which simplifies analysis.

%REFERENCES NEEDED FOR THIS JUNO OVERVIEW.
The only spacecraft currently operational at Jupiter, \textit{Juno}, has such an orbit.
%Juno is a Jupiter orbiter with a near-polar inclination 
%that was launched on 5 August 2011. Juno entered orbit around Jupiter on 4 July 2016.
Launched on 5 August 2011, \textit{Juno} entered orbit around 4 July 2016 and orbits with a near-polar inclination. 
The \textit{Juno} orbiter is the first spacecraft to operate in the outer solar system using solar power 
and the first to have a polar orbit around Jupiter \citep{2014Bagenal}. % an outer solar system object.
The nominal \textit{Juno} mission lifetime is 37 orbits. 
%NEED TO REWRITE BASED ON NEW LONGER ORBIT.
Four of these orbits are dedicated to spacecraft checkout and instrument commissioning, leaving 33 planned science orbits \citep{2016Connerney}.
\textit{Juno}'s perijove is at equatorial latitudes and 1.06 $R_J$, which is about 4300 km above the planet's cloud-tops.
Opportunities to conduct radio occultation observations of the IPT occur once per orbit.
Prior to orbit insertion, \textit{Juno} planned to conduct most of its mission in a 14-day orbit.
Due to anomalies encountered early in its orbital mission, the spacecraft may instead remain in a 53-day orbit for a considerable time.
The main findings of this article are not affected by the length of the orbital period as long as \textit{Juno} has a near-polar orbit with periapsis inside Io's orbit, which is true at the time of writing and likely to remain true until the end of the mission.
The only significant effect of changes from the planned orbital period is in temporal resolution. Measurements will be possible once per orbital period, so every 53 days instead of every 14 days. 

A major goal of the \textit{Juno} mission is to map Jupiter's gravitational and magnetic fields \citep{2014Bagenal}.
Analysis of these observations will improve understanding of the planet's interior structure and the properties of the magnetosphere out of the equatorial plane. 
The gravitational mapping requires continuous radio tracking from Earth, so the \textit{Juno} orbit is designed such that \textit{Juno} is never occulted from view by Jupiter itself. 
The \textit{Juno} project plans to conduct radio tracking on about 24 orbits \citep{2015Tommei}, which means that radio occultations may be feasible on those 24 orbits. 
With only two previous radio occultations of the IPT, the 24 possible \textit{Juno} occultations offer an order of magnitude increase in the number of observations and unprecedented opportunities to explore spatial and temporal variability in the IPT.
This set of occultations will sample the full range of System III longitudes and the full range of positions relative to Io along its orbit, but only a narrow range of local times. Due to the small angular separation of Earth and the Sun as seen from Jupiter, all occultations will be near noon local time.

The aims of this article are to evaluate the feasibility of measuring properties of the IPT with radio occultations conducted by the \textit{Juno} spacecraft, to estimate the likely accuracy of such observations, and to assess the contributions that such measurements could make towards key science questions concerning the IPT and its role in Jupiter's magnetosphere.
%This article also discusses the accuracy with which these observations can measure significant properties of the Io plasma torus. From these measurements we asses the contributions they could make towards better understanding the Io plasma torus and its role in Jupiter's magnetosphere. 

Section~\ref{sec:overview} describes the IPT.
Section~\ref{sec:occultation} discusses the concept of a radio occultation and relevant capabilities of \textit{Juno}.
%MAN, IT IS ANNOYING TO HAVE TO REMEMBER TO PUT ITALICS ON JUNO EVERY TIME. ANY REASON FOR THIS STYLISTIC DECISION?
Section~\ref{sec:analytic} explores radio occultations of the IPT using a simple model.
Section~\ref{sec:model} uses a more sophisticated model of the torus to determine the accuracy with which key torus properties can be measured.
Section~\ref{sec:discussion} discusses how plasma temperature and density can be obtained from the measured properties.
Section \ref{sec:conclusion} presents the conclusions of this work.

\section{\label{sec:overview}Overview of observations of the Io plasma torus}

The IPT can be observed in a variety of ways, including ground-based optical and infrared measurements \citep{1995Brown, 1995Schneider}, spacecraft in situ measurements \citep{1974Judge, 1975Carlson,1981Bagenal,1997Bagenal}, spacecraft ultraviolet (UV) measurements \citep{2004aSteffl,2004bSteffl}, and spacecraft radio occultation experiments \citep{1979Eshlemanb,1981Levy,1992Bird}.%HYPHEN IN GROUND-BASED HERE AND ELSEWHERE

Ground-based optical and infrared observations can measure the composition, density, and temperatures of plasma within the IPT \citep{1976Kupo, 1979Pilcher, 1995Brown, 1995Schneider}.
The intensities of species-specific emission lines indicate the composition of the plasma.
%These methods make measurements of species specific lines that are sensitive to the electron density and plasma temperature.}
%THE NEW SENTENCE ABOVE DOES NOT ACTUALLY MAKE SENSE.
%HOW DO THESE MEASUREMENTS PROVIDE COMPOSITION? PRESUMABLY FROM DETECTION OF SPECIES-SPECIFIC LINES AND RADIATIVE TRANSFER ANALYSIS. IF SO, SAY SO.
Electron density in the IPT can be determined from the intensity ratio of S$^{+}$ emission lines at 6717 and 6731 \AA~or the intensity ratio of O$^{+}$ emission lines at 3726 and 3729 \AA~ 
\citep{1976Brown}.
The electron temperature can be determined from intensity ratios of other pairs of S$^{+}$ lines and the
perpendicular ion temperature can be determined from the width of the S$^{+}$ 6731 \AA~line \citep{1976Brown}. %NEEDS TO BE SUPPORTED BY REFERENCE.
The brightest emissions from the IPT are sodium D-line emissions due to resonant scattering of solar radiation by neutral sodium, although their behavior is different from typical torus plasma since they are neutral. We can expect the sodium emission to be a tracer of the neutrals but not of the ions.
%The brightest emissions from the Io plasma torus are from scattered light from neutral sodium, even though sodium is a trace species. % OR IONIZED?
These emissions are often used as proxy measurements for the  main constituents of the IPT, ionized sulfur and oxygen.
Many ground-based surveys of spatial and temporal variability in the IPT have been conducted
\citep[e.g.][]{1995Brown, 1995Schneider, 2004Mendillob,2004Nozawa,2005Nozawa,2006Nozawa,2009Yoneda,2010Yoneda,2013Yoneda}. 

Valuable observations of the IPT were made by \textit{Voyager 1} during its flyby in March 1979 \citep{1981Bagenal} and \textit{Galileo} during its orbital tour in 1995--2003 \citep{1997Bagenal}. Each spacecraft was equipped with an ultraviolet spectrometer (UVS) that covered 400--1800 \AA~ 
and an in situ plasma instrument (PLS). 
The UVS experiments were able to measure electron density and temperature, ion temperature perpendicular to the magnetic field, and composition.
The PLS experiments were able to measure plasma density, velocity, and composition \citep{1994Bagenal,2004Thomas}. 
Due to degeneracies in the interpretation of observations from each instrument, 
both remote sensing UVS measurements and in situ PLS measurements were necessary to map the composition of the torus completely. 
In situ measurements by \textit{Voyager 1} \citep{1981Bagenal} and \textit{Galileo} \citep{1997Bagenal} mapped the spatial extent of the IPT. 
They found that the IPT is centered at the orbital distance of Io, 5.9 $R_{J}$, and 
has widths of about 2 $R_{J}$ in, and 1 $R_{J}$ perpendicular to, the plane of the centrifugal equator.
The Cassini Ultraviolet Imaging Spectrograph (UVIS) also observed the IPT during Cassini's Jupiter flyby in 2000--2001 \citep{2004aSteffl,2004bSteffl}.%NEED A REFERENCE TO SUPPORT THIS ALLEGATION.

The spatial distribution of plasma in the IPT has also been mapped by active remote sensing experiments on spacecraft in the Jupiter system. 
Prior to \textit{Juno}, radio occultations through the IPT have been conducted twice, once by \textit{Voyager 1} \citep{1979Eshlemanb,1981Levy,1985Campbell} and later by \textit{Ulysses} \citep{1992Bird,1993Bird}. 
These observations provided a time series of measurements of the TEC in the IPT between the spacecraft and Earth.
In combination with knowledge of the spacecraft trajectory, these TEC measurements constrained spatial variations in the local electron density within the IPT.   

%COMPOSITION
%DENSITY
%SPATIAL EXTENT
%VELOCITY?
%TEMPERATURE
%EQUATION THAT LINKS PLASMA SCALE HEIGHT TO TEMPERATURE
%ANYTHING ELSE INTERESTING
From these observations, a general picture of the IPT has been developed. From \textit{Voyager 1} measurements,
\citet{1981Bagenal} found that the torus can be divided into three different regions: the cold torus, ribbon, and warm torus. 
The innermost region, centered at 5.2 $R_{J}$, is the cold torus. %The center of the cold torus is at a distance of 5.2 $R_{J}$.
In the cold torus, densities fall off with height above the centrifugal equator with a scale height of 0.1 $R_{J}$, which is relatively small.
The cold torus peaks at around 5.23 $R_{J}$ and extends from 4.9 $R_{J}$ to 5.5 $R_{J}$ and has a characteristic density of 
$\sim$1000 cm$^{-3}$. Its composition is mostly S$^+$ ions with smaller amounts of O$^{+}$ ions present.
In the cold torus, the electron temperature $T_{e} \approx$ 1--2 eV and the ion temperature $T_i \approx$ 1--4 eV. 
%MUST DISCUSS EQN 10 IN THIS CONTEXT of 0.6 RJ.
Beyond the cold torus lies the ribbon, whose center is at a distance of 5.6 $R_{J}$.
It has a scale height of 0.6 $R_{J}$ and extends from 5.5--5.7 $R_{J}$.
The ribbon has a high characteristic density of $\sim$3000 cm$^{-3}$ and it is mostly O$^+$ ions with smaller amounts of S$^{+}$ ions present.
In the ribbon, $T_e\approx$ 4--5 eV and $T_i \approx$ 10--30 eV.
The outermost region is the warm torus, whose center is at Io's orbital distance of 5.9 $R_{J}$.
It has a scale height of 1 $R_J$, which makes it the thickest region, and extends from 5.7--8 $R_{J}$.
The warm torus has a characteristic density of $\sim$2000 cm$^{-3}$ and it is composed of S$^{2+}$ and O$^{+}$ ions 
with trace amounts of O$^{2+}$, S$^{+}$, and S$^{3+}$ ions. 
In the warm torus, $T_e \approx$ 5--8 eV and $T_i \approx$ 60 eV.
%I REALLY NEED TO FIGURE OUT HOW A WIDTH OF 2 RJ AGREES WITH RESTRICTION IN EQN 10 TO 5.7 TO 6.1 RJ. VERY CONFUSING.

The scale height, $H$, is related to plasma composition and temperature
\citep{1992Thomas,2004Thomas},
%XXX4 - We should remember to work out the derivation of this equation

\begin{equation}
H = \sqrt{\frac{2k(T_{i,\parallel}+Z_{i}T_{e,\parallel})}{3M_{i}\Omega^2}}
\label{eqn:scaleheight}
\end{equation}
where $k$ is the Boltzmann constant, $T_{i,\parallel}$ is the ion temperature, $Z_{i}$ is the atomic number of the ion species,
$T_{e,\parallel}$ is the electron temperature, $M_{i}$ is the mass of the ion species, and $\Omega$ is the rotation rate of Jupiter's magnetosphere ($\sim 1.75 \times 10^{-4}$ rad s$^{-1}$). %IS THIS WHAT ANGULAR VELOCITY MEANS?
The $\parallel$ subscripts on $T_{e}$ and $T_{i}$ refer to the component of temperature parallel to the magnetic field.
Since $T_{e}$ is much smaller than $T_{i}$, the scale height is effectively insensitive to $T_{e}$.
%DO YOU KNOW WHY TE IS MUCH SMALLER THAN TI?
This scale height defines the extent of the IPT parallel to the magnetic field lines. % normal to the centrifugal equatorial plane
If a radio occultation can determine the scale height $H$, then Equation~\ref{eqn:scaleheight} can be used to infer the ion temperature.
Doing so requires independent knowledge of the ion composition, which is summarized above.
%INCORPORATE THIS BIT INTO THE ABOVE SUMMARY OF THE TORUS
%\citet{1981Bagenal} found from the \textit{Voyager 1} measurements that the torus is broken into three different regions: the cold torus ($T_e\approx$ 1--4 eV, $T_i \approx$ 10 eV), the ribbon($T_e\approx$ 4--5 eV, $T_i \approx$ 50 eV ), and the warm torus ($T_e\approx$ 5--8 eV, $T_i \approx$ 70 eV). 

Previous observations have revealed much about the plasma torus. 
However, many key science questions still remain concerning the generation, transport, and loss of plasma in the IPT and the magnetosphere of Jupiter. 
%As noted in Section \ref{sec:intro}, %XXX INTRODUCTION XXX as noted in Section \ref{sec:intro},
%the generation, transport, and loss of plasma associated with the Io plasma torus are the subjects of key science questions about the magnetosphere.
In the context of the IPT itself, outstanding questions include:
1. Over what timescales does the supply of plasma to the IPT vary?
2. How do variations in Io's volcanic activity affect major properties of the IPT?
3. How do major properties of the IPT vary with System III longitude?
Multiple radio occultations of the IPT by \textit{Juno} will provide new information for answering these questions.
These radio occultations offer unparalleled spatial and temporal coverage of the IPT.

\section{\label{sec:occultation}Radio occultations}

A radio occultation occurs when an object, here the IPT, comes between the transmitter and receiver of a radio signal.
On each orbit, \textit{Juno} will pass through the centrifugal equator such that the IPT is between the spacecraft and Earth. 
This geometry is suitable for radio occultation observations of the torus.

%I ASSUME JUNO USES TWO-WAY RADIO LINK, BUT WE NEED TO VERIFY THAT
During radio tracking, \textit{Juno} will receive a radio signal from Earth at X-band frequencies (7.3 GHz) and use multipliers to retransmit that signal back to Earth at X-band frequencies (8.4 GHz) and Ka-band frequencies (32.1 GHz) (Table \ref{tab:Junoparamtable}) \citep{2012Mukai}. 
This method is similar to the method used by Cassini for radio occultations after the failure of its ultrastable oscillator \citep{2015Schinder}
and to the method that will be used by \textit{BepiColombo} for gravity science  measurements \citep{2015Tommei}. 
%,2001Iess}.}NEED A REFERENCE HERE, SCHINDER RADIO SCIENCE 2015 OUGHT TO BE OK  SOME KIND OF SCIENCE 
Since the downlink frequencies are derived from the same source, the two down-linked radio signals will be transmitted coherently.

The propagation of the radio signal is affected by plasma along its path such that 
the received frequency contains information about the electron density along the path of the radio signal.
As is shown here, 
the line-of-sight integrated electron density can be derived from comparison of the received frequencies of the two down-linked radio signals.

%CITE SOURCE FOR EQN
Neglecting relativistic effects, the received frequency on Earth of the downlinked X-band signal satisfies \citep{2014Withers}:

\begin{eqnarray}
f_{R,X} = f_{T,X} - \frac{f_{T,X}}{c} \frac{d}{dt} \int{}dl + 
\frac{e^2}{8 \pi^2 m_e \epsilon_0 c f_{T,X}} \frac{d}{dt} \int{} N dl -\frac{f_{T,X}\kappa}{c} \frac{d}{dt} \int{} n dl
\label{eqn:X_band}
\end{eqnarray}
where $f$ is frequency, subscripts $R$ and $T$ refer to received and transmitted, respectively, subscript $X$ refers to X-band, $c$ is the speed of light, $t$ is time, $l$ is distance along the ray path, $-e$ is the electron charge, $m_e$ is the electron mass, $\epsilon_0$ is the permittivity of free space, $N$ is the electron density, $\kappa$ is the mean refractive volume of the neutrals, and $n$ is the number density of neutrals.
A similar equation can be written for the received frequency on Earth of the downlinked Ka-band signal:

\begin{eqnarray}
f_{R,Ka} = f_{T,Ka} - \frac{f_{T,Ka}}{c} \frac{d}{dt} \int{}dl + 
\frac{e^2}{8 \pi^2 m_e \epsilon_0 c f_{T,Ka}} \frac{d}{dt} \int{} N dl -  \frac{f_{T,Ka}\kappa}{c} \frac{d}{dt} \int{} n dl
\label{eqn:Ka_band}
\end{eqnarray}
The two transmitted frequencies, $f_{T,X}$ and $f_{T,Ka}$, satisfy $f_{T,Ka} / f_{T,X} = f_{D,Ka} / f_{D,X}$, where $f_{D,Ka} / f_{D,X}$ is a fixed ratio of 3344/880 \citep{2004kliore}.
% This is supported by Fig 11 of Kliore 2004
%XXX5 - Need reference to support 336/88 statement. 
The subscript D refers to downlinked frequencies.
Accordingly, Equation~\ref{eqn:Ka_band} can be multiplied by $f_{D,X} / f_{D,Ka}$ and subtracted from Equation~\ref{eqn:X_band} to give:

\begin{eqnarray}
\Delta f = f_{R,X}-f_{R,Ka} \left( \frac{f_{D,X}}{f_{D,Ka}} \right) = 
\frac{e^2}{8 \pi^2 m_e \epsilon_0 c f_{T,X}} \left( 1- \left( \frac{f_{D,X}}{f_{D,Ka}} \right)^2 \right) \frac{d}{dt} \int{} N dl
\label{eqn:shiftfreq}
\end{eqnarray}
where $\Delta f$ is defined as $f_{R,X}-f_{R,Ka} \left( \frac{f_{D,X}}{f_{D,Ka}} \right)$.
Terms proportional to the transmitted frequency in Equations~\ref{eqn:X_band}--\ref{eqn:Ka_band} cancel out in this difference.
This eliminates the classical Doppler shift and effects of neutral molecules.
The quantity $\int{} N dl$ is the line-of-sight TEC.
If time series of $f_{R,X}$ and $f_{R,Ka}$ are available, Equation~\ref{eqn:shiftfreq} can be used to determine the rate of change of the TEC.
Given knowledge of the spacecraft trajectory, the time rate of change of the TEC can be converted into the spatial gradient of the TEC.
Finally, this can be integrated to give the TEC for each different line of sight.

%A quick back of the envelope calculation can be done to determine the how the shifted frequency would be affected by changes in the scale height and peak plasma density during an occultation of the \textit{Juno} spacecraft. 

\section{\label{sec:analytic}Initial estimate of frequency shifts}

We wish to determine how accurately properties of the IPT can be measured by radio occultation experiments.
Before developing a sophisticated model of the IPT and sources of noise, we first explore the influence of IPT properties on observable quantities using a simple model.

\subsection{\label{sec:toyiptmodel}Initial model of Io plasma torus}

%NEED TO BE CAREFUL WITH SYMBOLS HERE.
%LATER YOU HAVE BIG-R AND SMALL-R AS DISTINCT QUANTITIES.
%HERE WE WANT CYLINDRICAL SYMMETRY, SO CAN'T USE EITHER BIG-R OR LITTLE-R
%We assume that electron density $N$ depends on distance from the center of the torus $s$ is represented by a single Gaussian function as follows 
We assume that the electron density $N$ can be represented by a single Gaussian that depends on the distance $s'$ from the center of the torus 
and that the center of the torus lies in the plane of the centrifugal equator at a distance from Jupiter equal to Io's orbital distance of 5.9 $R_{J}$
\citep{1992Thomas,2004Thomas}. 
%\textbf{Interactions of the plasma with the magnetic field of Jupiter can cause deviations from this distribution but we will ignore those dependencies in this paper.}
%THIS ISN'T QUITE CLEAR YET, LET'S DISCUSS.
%THE LINE ABOUT MAGNETIC FIELD DEPENDENCIES GOES HERE - ONCE ITS MEANING IS EXPRESSED CLEARLY.
% \textbf{ignoring any magnetic field line dependencies}.
Hence:

% PROBLEM - IN THESE EQUATIONS, IS E E-CHARGE OR E-BASEOFLOGS? NO GOOD SOLUTION EXISTS
\begin{equation}
N\left(s'\right) = N\left(0\right)\exp^{-\frac{s'^2}{H^2}}
\label{eqn:distribution}
\end{equation}
%LET'S DISCUSS THAT SQRT 2 FACTOR AND DEFINITION OF SCALE HEIGHT. FIND SOME SOURCES WHERE OTHER WORKERS HAVE USED A SCALE HEIGHT TO DESCRIBE THE PLASMA TORUS AND FIGURE OUT WHAT THEIR DEFINITIONS ARE.
%NEED TO CITE SOME SOURCES FOR THIS FUNCTIONAL FORM BEING REASONABLE.
%AT SOME POINT, NEED TO STATE A SINGLE REPRESENTATIVE VALUE FOR N(0) AND A SINGLE REPRESENTATIVE VALUE FOR H.electron column density (or total electron content)
where $H$ is the scale height and $N\left(0\right)$ is the density at the center of the torus. Typical values for $N\left(0\right)$ and $H$ are 2000 cm$^{-3}$ and 1 $R_J$, respectively \citep{2004Thomas}. %CITE SOURCE.
The critical quantity in Equation~\ref{eqn:shiftfreq} is $\int N dl$, the integral of the electron density along the line of sight.
We define TEC as a function of the radio signal's distance of closest approach to the center of the torus $s$, $TEC\left(s\right)$.
This satisfies \citep{2006Quemerais}:

\begin{equation}
\int N dl = TEC\left(s\right) = 2\int_{s}^{\infty} \frac{N\left(s'\right)s' ds'}{\sqrt{s'^2-s^2}}
\label{eqn:integral}
\end{equation}
%NO, THIS IS NOT WHAT THE DUMMY VARIABLE R (NOW CHANGED TO S') MEANS IN THIS EQUATION, DRAW A DIAGRAM TO SEE WHAT IT ACTUALLY IS. SINCE YOU ARE INTEGRATING OVER IT, IT CAN'T BE A SINGLE-VALUED THING LIKE THE POSITION TO POINT OF CLOSEST APPROACH.
where s' is the distance from the center of the IPT to a point on the ray path.
%\textbf{where R is the distance from the center of Jupiter to the point on the ray path where the ray reaches closest approach distance $s$.} 
With the density $N$ given by Equation \ref{eqn:distribution}, $TEC\left(s\right)$ is given by \citep{1972Abramowitz}:
%XXX6 - What's the source of this result? Is the integral easy? Or from a reference book?

\begin{equation}
TEC\left(s\right) = 
N(0) \sqrt{\pi} H \exp^{-\frac{s^2}{H^2}}
\label{eqn:coldenssoln}
\end{equation}
The maximum value of $TEC\left(s\right)$ occurs at $s=0$, where $TEC = N(0)\sqrt{\pi}H$. 
For benchmark values $N\left(0\right)=2000$ cm$^{-3}$ and $H=1$ $R_J$, the maximum value of the TEC is $25.5 \times 10^{16}$ m$^{-2}$.
This can be expressed as 25.5 TECU, where 1 TECU or total electron content unit equals $1 \times 10^{16}$ m$^{-2}$.
%HAVE I GOT THE NUMERICAL FACTORS RIGHT IN THE SOLUTION? DOES METHOD OF OBTAINING THIS SOLUTION NEED A REFERENCE TO A BOOK OF INTEGRALS OR SOMETHING SIMILAR? I FORGET IF THE SOLUTION IS EASY OR NOT.

\subsection{\label{sec:charfreqshift}Characteristic frequency shifts}

Combining Equations \ref{eqn:shiftfreq} and \ref{eqn:coldenssoln}, the frequency shift $\Delta f$ satisfies:

%LEAVE THE CONSTANTS, GET RID OF THE NUMERICAL VALUE. EQUATION IS NOT DIMENSIONALLY CORRECT AS WRITTEN.
\begin{eqnarray}
\Delta f\left(s\right) = \frac{e^2}{8 \pi^2 m_e \epsilon_0 c f_{T,X}}\left( 1- \left( \frac{f_{D,X}}{f_{D,Ka}} \right)^2 \right) \frac{d}{dt} \left[N\left(0\right) \sqrt{\pi} H \exp^{-\frac{s^2}{H^2}} \right]%1.498x10^{-17}
\label{eqn:deltaf}
\end{eqnarray}
Since the only time-variable quantity in Equation \ref{eqn:coldenssoln} is the distance of closest approach $s$, Equation \ref{eqn:deltaf} becomes:

%LEAVE THE CONSTANTS, GET RID OF THE NUMERICAL VALUE. EQUATION IS NOT DIMENSIONALLY CORRECT AS WRITTEN.
\begin{eqnarray}
\Delta f\left(s\right) = -\frac{e^2}{8 \pi^2 m_e \epsilon_0 c f_{T,X}} \left( 1- \left( \frac{f_{D,X}}{f_{D,Ka}} \right)^2 \right) \sqrt{\pi} N\left(0\right) \exp^{-\frac{s^2}{H^2}} \left(\frac{2s}{H}\right) \frac{ds}{dt}
\label{eqn:analytic}
\end{eqnarray}
Here $\frac{ds}{dt}$ is the rate of change of the distance of closest approach $s$. 
Note that this refers to the distance of closest approach of the line of sight between the spacecraft and Earth to the center of the IPT.
It is therefore affected by the trajectory of the spacecraft and the motion of the IPT, not solely by the trajectory of the spacecraft.
For simplicity in this exploratory work, we assume that $ds/dt$ is constant during a radio occultation observation.
However, this is a questionable assumption that would need to be revised in the analysis of real observations.
First, \textit{Juno}'s speed during a radio occultation observation, which is essentially a periapsis pass, changes appreciably due to the high eccentricity of \textit{Juno}'s orbit.
Second, since the IPT is tilted with respect to Jupiter's rotational axis, the center of the IPT moves during a radio occultation observation.
At Io's orbital distance, the center of the IPT moves up and down with a velocity of $\pm$ 9 kms$^{-1}$ over Jupiter's 9.925 hour rotational period.
%PLEASE FILL IN THE XXX
We assume that $\left|ds/dt\right|$ is 20 km s$^{-1}$, which is a representative value for the spacecraft speed during a periapsis pass
(based on the ephemeris tool at www-pw.physics.uiowa.edu/$\sim$jbg/juno.html).
This is equivalent to a change in $s$ of one R$_J$ in a time of one hour. 
We reconsider this issue at the end of Section~\ref{sec:charfreqshift}.

%It is not the radial speed of the spacecraft.
%Instead, it is the component of the velocity of the spacecraft in the direction normal to the centrifugal equatorial plane.%$z$-direction. 
%HAVE YOU DEFINED Z? OR A COORDINATE SYSTEM IN GENERAL, IF SUCH IS NEEDED? I ASSUME Z IS POSITION PERPENDICULAR TO PLANE OF THE TORUS.%-5.313x10^{-17}

%THERE'S AN ANNOYING ISSUE HERE. WE DEFINE S (PREVIOUSLY R) AS CLOSEST APPROACH DISTANCE, WHICH MEANS IT IS POSITIVE DEFINITE.
%YET IT IS CLEARLY CONVENIENT TO USE +/- VALUES TO DISTINGUISH BETWEEN INGRESS AND EGRESS PARTS OF OCCULTATION. 
%I DON'T KNOW WHAT THE SOLUTION IS, LET'S WORRY ABOUT IT LATER. 

Equation \ref{eqn:analytic} provides an analytical description of the dependence of the measureable frequency shift $\Delta f$ on the central density of the torus, $N\left(0\right)$, the torus scale height, $H$, and $ds/dt$, 
which can be intepreted as the projected speed of the spacecraft.
The value and location of the maximum value of $\left|\Delta f\right|$ can be found by setting the derivative of Equation~\ref{eqn:analytic} with respect to $s$ to zero.
The maximum value of $\left| \Delta f\right|$, $\left| \Delta f\right|_{max}$, occurs at $s^{2} = H^{2} / 2$ and satisfies:

\begin{equation}
\left| \Delta f\right|_{max} =
\frac{e^2}{8 \pi^2 m_e \epsilon_0 c f_{T,X}} \left( 1- \left( \frac{f_{D,X}}{f_{D,Ka}} \right)^2 \right) \sqrt{\pi} N\left(0\right) e^{-\frac{1}{2}} \sqrt{2} \frac{ds}{dt}
\label{max}
\end{equation}

For $N\left(0\right)$ = 2000 cm$^{-3}$, $H$ = 1 $R_J$, and $\left|ds/dt\right|$ = 20 km s$^{-1}$, the maximum value of $\left| \Delta f\right|$ is 0.9 mHz.
This maximum occurs at $s = 0.7 R_{J}$. 
% which is the average velocity of the spacecraft at closest approach calculated with the University of Iowa ephemeris tool.   

\begin{figure}
\noindent\includegraphics[width=20pc]{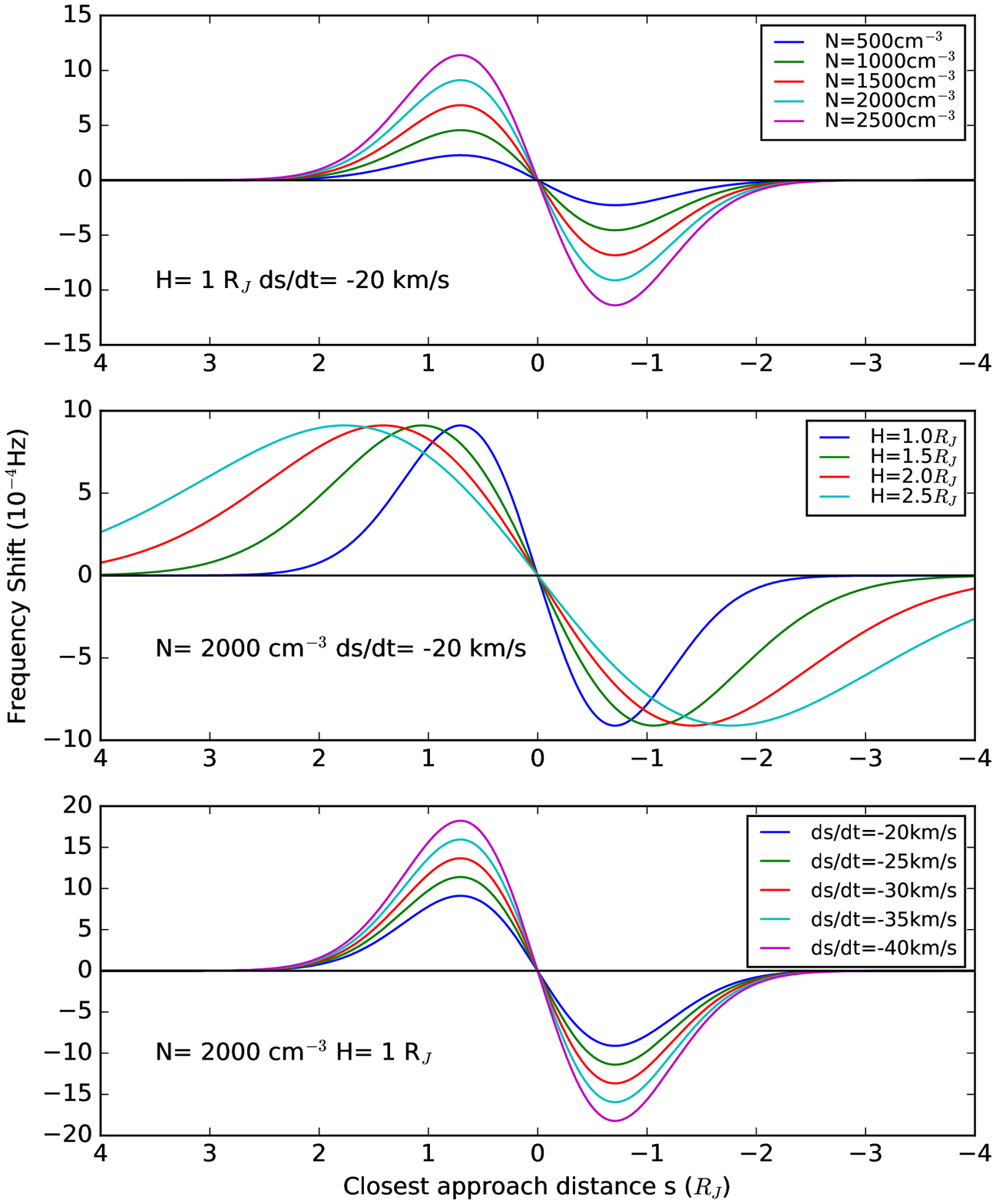}
\caption{(top) Dependence of frequency shift on closest approach distance $s$ for initial model of torus. Different lines represent different values for $N\left(0\right)$.
(middle) Dependence of frequency shift on closest approach distance $s$ for initial model of torus. Different lines represent different values for $H$.
(bottom) Dependence of frequency shift on closest approach distance $s$ for initial model of torus. Different lines represent different values for $ds/dt$.
The quantities that are held fixed in each panel are shown in the bottom left corners.
%XXX17 - Change xtitle minimal distance to closest approach distance. Also figures 2 and 3.
}
%This figure shows how the shift in frequency is affected by different peak densities. Since r is a function of time the x-axis can be thought of as time increasing to the right.}
\label{fig:test}
\end{figure}

The top panel of Figure \ref{fig:test} shows how $\Delta f$ depends on $s$ for several values of $N\left(0\right)$ and fixed $H$ = 1 $R_J$ and $ds/dt$ = -20 km s$^{-1}$. 
We choose a range of values for $N\left(0\right)$ that covers the observed values in the torus.
$N\left(0\right)$ varies between 500 cm$^{-3}$ and 2500 cm$^{-3}$ \citep{1981Bagenal, 1997Bagenal}.
The middle panel of Figure \ref{fig:test} shows how $\Delta f$ depends on $s$ for several values of $H$ and fixed $N\left(0\right)$ = 2000 cm$^{-3}$ and $ds/dt$ = -20 km s$^{-1}$.
We choose a range of values for $H$ that covers the observed values in the torus.
$H$ varies between 0.5 R$_J$ and 2.5 R$_J$ \citep{2004Thomas}.
The bottom panel in Figure \ref{fig:test} shows how $\Delta f$ depends on $s$ for several values of $\left|ds/dt\right|$ and fixed $N\left(0\right)$ = 2000 cm$^{-3}$ and $H$ = 1 $R_J$.
We choose a range of values for $\left|ds/dt\right|$ that increases from 20 km s$^{-1}$ to 40 km s$^{-1}$ in increments of 5 km s$^{-1}$. 

%XXX7 - What is maximum value of Deltaf and at what value of s (formula, not number) does it occur? Differentiate Eqn 9 with respect to s, set equal to zero, solve. I find it interesting that Deltaf-max is independent of H.~\ref{fig:Ntest}--\ref{fig:Vtest}
Figure \ref{fig:test} illustrates how the observed shift in frequency, $\Delta f$, depends on $N\left(0\right)$, $H$, and $ds/dt$.
$\Delta f$ is zero at the start of an occultation, when $\left|s\right|$ is large.
Its magnitude increases monotonically to $\Delta f_{max}=$ 0.9 mHz at $s_{crit} = H/\sqrt{2}$
%THIS IS DIMENSIONALLY INCORRECT, NEEDS TO BE FIXED
, then decreases monotonically through zero at $s=0$.
The behavior of $\Delta f$ in the second half of the occultation is the same as in the first half, except for a change in sign.
The full width at half maximum of the local maximum in $\Delta f$ is approximately equal to $H$.

The effects of variations in $N\left(0\right)$ and $ds/dt$ are straight-forward, since the frequency shift $\Delta f$ is proportional to both factors.
Spatial and temporal changes in $N\left(0\right)$ are likely over the course of the \textit{Juno} mission, since the IPT is intrinsically variable, whereas $ds/dt$ will not vary greatly from orbit to orbit.
%be relatively constant, like the \textit{Juno} orbit.
The effects of variations in $H$ are more complex.
As $H$ increases, $s_{crit}$ increases. The width of the local maximum in $\Delta f$ also increases, but the value of $\Delta f_{max}$ remains the same.

%Since equation \ref{eqn:analytic} is proportional to both N0 and  $ds/dt$ the affect that they have are exactly the same. The \textit{Juno} spacecraft will be traveling at relatively the same velocity over multilple orbits and thus N0 will be the only quantity changing from orbit to orbit. Changes in H cause the observations to change in radial with of the peak. Thus the measured widths of the peak are proportional to the value of H. 
%HAVING SHOWN THESE FIGURES, NOW MUST PROVIDE SOME INTERPRETATION OF THEM HERE...

The timescale, $\tau$, for the radio signal to sweep through the IPT satisfies $\left|ds/dt\right| \tau = 2 H$.
With $\left|ds/dt\right|= 20$ km s$^{-1}$ and $H = 1 R_J$, the timescale $\tau$ is approximately 2 hours.

%SAY SOMETHING ABOUT TYPICAL SPEED, TYPICAL H, TIME FOR DR/DT X DT TO EQUAL 2 H. THIS TELLS READER HOW LONG OBSERVATION WILL TAKE.

An integration time on the order of 10 seconds provides spatial resolution on the order of $H/100$.
For this integration time, it can be assumed that
the relative accuracy with which $f_{R,X}$ and $f_{R,Ka}$ can be measured is $3 \times 10^{-14}$.
This is based on the Allan deviation of the Deep Space Network (DSN) hydrogen masers over a 10 second integration \citep{1992Howard, 2005Asmar}.
%REF HOWARD 1992.
%For typical values of H and $dr/dt$ give an integration time of 36 seconds to reach 1/100 of H. Thus with an integration time of 10 seconds we can get a resolution of 0.4 percent of H. 
%FOR TYPICAL VALUES OF H AND DR/DT, WHAT INTEGRATION TIME IS NEEDED TO GIVE RESOLUTION OF 1/100 OF H? IS 10 SECOND ABOUT RIGHT? THIS IS WORTH DISCUSSING IN TEXT.
%I PRESUME WE KNOW THAT JUNO CARRIES A USO, WOULD BE NICE TO CITE AND STATE THAT.
%JUNO USO PERFORMANCE PROBABLY CLOSER TO CASSINI USO DUE TO DEVELOPMENTS IN USO TECHNOLOGY OVER TIME, HAVE A LOOK AT RELEVANT KLIORE SSR ARTICLE.
With $f_{R,X}$ = 8.4 GHz and $f_{R,Ka}$ = 32.1 GHz \citep{2012Mukai}, the corresponding uncertainty in a measurement of $\Delta f$ is $3.8 \times 10^{-4}$  Hz (Equation~\ref{eqn:shiftfreq}).
% CITE SOURCE AND THINK ABOUT HOW MANY DECIMAL PLACES ARE RELEVANT HERETHINK ABOUT HOW MANY DECIMAL PLACES ARE RELEVANT HERE. 
This uncertainty in $\Delta f$, $\sigma_{\Delta f}$, is 40 percent of the characteristic value of 0.9 mHz discussed above. 
%XXX8 - No, it isn't. Some value is written wrongly.

The uncertainty on the inferred TEC, $\sigma_{TEC}$, follows from propagating the uncertainty in $\Delta f$ through the integrated version of Equation~\ref{eqn:shiftfreq}.
Assuming a simple numerical integration method leads to:
\begin{equation}
\sigma_{TEC} = \sqrt{\Sigma}\left(\frac{e^2}{8 \pi^2 m_e \epsilon_0 c f_{T,X}} \left( 1- \left( \frac{f_{D,X}}{f_{D,Ka}} \right)^2\right)\right)^{-1}\sigma_{\Delta f}\Delta t
\label{eqn:sqrtsigma}
\end{equation}
where $\Sigma$ is the number of data points integrated to reach the current measurement and $\Delta t$ is the integration time for an individual measurement. 
Since $\Sigma = t/\Delta t$, where $t$ is the time since the start of the observation,
we obtain:
\begin{equation}
\left(\frac{\sigma_{TEC}}{1 \: \mathrm{TECU}} \right) = 0.5 %0.487 
\sqrt{\left(\frac{t}{1 \: \mathrm{hr}}\right) \left(\frac{\Delta t}{10 \: \mathrm{s}} \right)}
\label{eqn:uncertainty}
\end{equation}

%XXX9 - Something very, very wrong in this equation. Does 4.87E15 belong here?
For $N\left(0\right)$ = 2000 cm$^{-3}$, $H$ = 1 $R_J$, and $\left|ds/dt\right|$ = 20 km s$^{-1}$, $\Delta f_{max}$ = 0.9 mHz and $s_{crit}$ = 0.7 $R_{J}$.
If the integration starts at $s$ = 4 $R_J$, then $t$ at this local maximum is 3.3 hours from the start of the observation.
Henceforth we adopt $\Delta t$ = 36 seconds to provide a resolution of 0.01 $R_{J}$.
This yields $\sigma{TEC} / \left(1 \: \mathrm{TECU} \right) = 0.92 \sqrt{ t / \left(1 \mathrm{hr} \right)}$, which gives $\sigma_{TEC}$ = 1.68 TECU at the local maximum.
In this example, $\sigma_{TEC} / TEC$ = 7\% at the TEC maximum.

Several other potential sources of error must be considered.
The effects of noise at the transmitter and receiver on the simulated measurements of frequency shift are accounted for by the stated Allan deviation.
The effects of plasma in the rest of Jupiter's environment and the interplanetary medium can be accounted for in the frequency baseline prior to and after the occultation of the IPT \citep[e.g][]{2000Thornton}. 
The noise contribution due to the interplanetary medium depends strongly on solar elongation angle. It should be noted that for most of the \textit{Juno} mission the solar elongation angle is relatively large and the associated noise is relatively small \citep{1979Woo, 2005Asmar}.
%AND THIS IS GOOD/BAD, AS SUPPORTED BY SOME SOURCE...
%THIS WOO AND ARMSTRONG PAPER IS OK, THEIR FIG 3 - http://adsabs.harvard.edu/abs/1979JGR....84.7288W
Plasma in the regions of Jupiter's magnetosphere outside the IPT will also contribute to the measured TEC.
At the centrifugal equator, assuming a magnetospheric density 3 cm$^{-3}$ and length of 100 $R_{J}$ \citep{2015Bolton}, this contribution is about 0.7 TECU, which is small (3\%)~ relative to the peak TEC of the IPT, 25.5 TECU.

\textit{Juno}'s periapsis altitude is approximately 4000 km, which is within the ionosphere \citep{2014Bagenal}. Hence plasma in Jupiter's ionosphere may contribute to the measured total electron content between the spacecraft and Earth. The ionospheric plasma density at this altitude is approximately $3\times 10^{9}$ m$^{-3}$ and the ionospheric scale height is on the order of 1000 km \citep{1979Eshlemana,2004Yelle}. This results in a vertical total electron content of $3 \times 10^{15}$ m$^{-2}$ or 0.3 TECU. The line of sight total electron content will be larger by a geometric factor.  This is a potentially significant perturbation to the inferred total electron content of the IPT, especially if passage through the ionosphere occurs as the line of sight to Earth passes through the centrifugal equator. However, the \textit{Juno} Waves instrument is capable of measuring the local plasma density at the spacecraft \citep{2014Bagenal}. Using its measurements of the vertical structure of the topside ionosphere, the contributions of Jupiter's ionosphere to the inferred total electron content of the IPT can be eliminated. 
%THERE'S A GREAT BIG ASSUMPTION ABOUT THE LENGTH OF THE MAGNETOSPHERE IF YOU ARE GOING TO LEAP FROM 3 CC TO 0.7 TECU. THE READER MIGHT LIKE TO BE TOLD THIS LENGTH. ALSO REFERENCES TO SUPPORT YOUR CLAIMED NUMBERS.

Since Io orbits Jupiter every 1.7 days, each occultation will measure IPT properties at a different angular separation from Io. A series of occultations over a range of separations from Io will be valuable for assessing how plasma is transported away from Io and into the IPT. It is possible, though unlikely, for an IPT occultation to also probe Io's ionosphere directly. In that event, the line-of-sight total electron content would briefly increase by 0.1 TECU or $1 \times 10^{15}$ m$^{-2}$. This follows from a surface ionospheric density of $6 \times 10^{3}$ cm$^{-3}$ and a scale height of 100 km \citep{1998Hinson}.
%{Io orbits Jupiter ever approximately 1.7 days. Thus, the position of Io relative to the occultation point is important for assessing the transport of plasma away from Io. The orbit allows for there to be a chance for occultation of Io's ionosphere. With a ionospheric density of $10 \times 10^{3}$ cm$^{-3}$ and a scale height of 100 km (\citep{1998Hinson} the IPT profile will increase by approximately 0.1 TECU, or $1 \times 10^{15}$ m$^{-2}$.}

\begin{figure}
\noindent\includegraphics[width=20pc]{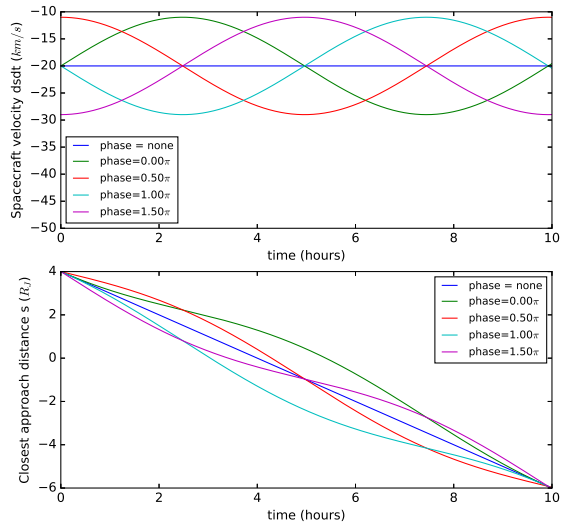}
\caption{(top) 
Dependence of $ds/dt$ on time when the spacecraft is moving at a constant speed of 20 km s$^{-1}$ and the Io plasma torus is moving with a 
sinusoidally-varying speed that has a period of 9.925 hours and an amplitude of 9 km s$^{-1}$.
Different lines show different phases for the motion of the Io plasma torus. The horizontal line that corresponds to a fixed Io plasma torus is for reference.
%TEXT SAID 9.9 KM/S, WHICH I ROUNDED TO 10 KM/S, THIS CAPTION PREVIOUSLY SAID 9. PICK A NUMBER, JUSTIFY IT ON FIRST USE, AND BE CONSISTENT WITH IT.
%%FIGURE NEEDS TO BE REVISED. THE TIME AXIS HAS NO GOOD REASON FOR STARTING AT 6 HRS, MIGHT AS WELL START AT ZERO. ALSO NEED TO JUSTIFY THE LENGTH OF TIME SHOWN. LATER PLOTS HAVE 10 HOUR DURATION, MIGHT AS WELL ADOPT THAT HERE. BOTH PANELS MUST HAVE SAME X-AXES, THEY CURRENTLY DO NOT.
%%FOR CLARITY OF PLOT, NEED TO HAVE SINES=0 AND COSINES=+/-1 AT THE LEFT-HAND AXIS. THE CURRENT PLOT STARTS ALL SINUSOIDS AT AN UTTERLY RANDOM PHASE.
(bottom) 
Corresponding dependence of distance of closest approach, $s$, on time.
Line colors are as in the top panel. The diagonal line that corresponds to a fixed Io plasma torus and constant $ds/dt$ is for reference.
}
\label{fig:test_v_shift}
\end{figure}

We previously noted the flaws in the assumption that $\frac{ds}{dt}$ is constant.
There are two main consequences if $\frac{ds}{dt}$ is not constant.
The first consequence is that it becomes harder to determine the position $s$ associated with a given time in the measured time series of $\Delta f$.
Yet since the \textit{Juno} trajectory and the location of the centrifugal equator at Io's orbital distance are known, the required mapping from time to position is tractable.
The effects of the nodding up and down of the IPT are illustrated in Figure \ref{fig:test_v_shift}.
This shows how $\frac{ds}{dt}$ and $s\left(t\right)$ change for different phasings of the motion of the IPT relative to the time of the occultation. This is equivalent to occultations occurring at different System III longitudes.
From a fixed vantage point of noon local time in the rotational equator, the IPT moves up and down sinusoidally with a period equal to the planetary rotation period of 9.925 hours, a distance magnitude of 5.89 $R_{J}$, and a speed magnitude of 9 km s$^{-1}$.
Given a constant spacecraft speed of 20 km s$^{-1}$, which is itself a noteworthy simplification, $\left|ds/dt\right|$ varies between 10 and 30 km s$^{-1}$.
The variation in $ds/dt$ with time leads to the second consequence, which is that the numerical and graphical results based on Equations \ref{eqn:deltaf} and \ref{eqn:analytic} will no longer be perfectly accurate. %XXX PUT PROPER CROSS-REFERENCES IN FOR EQNS 9 AND 10 XXX
Furthermore, note that a constant time resolution in the measured received frequencies will no longer correspond to a constant spatial resolution within the IPT.
%I AM OK WITH SWEEPING THE IPM UNDER THE RUG IN THIS WAY, BUT WE NEED TO THINK ABOUT JUPITER'S MAGNETOSPHERE.
%THE MAGNETOSPHERIC CONTRIBUTION TO THE TEC WILL CHANGE THROUGH THE OCCULTATION. NEED TO BACK-OF-THE-ENVELOPE IT WITH MANY SIMPLIFYING ASSUMPTIONS.
%I AM CURIOUS TO CONSIDER THE LIKELY EFFECTS OF SWEEPING THROUGH JUPITER'S MAGNETOSPHERE OVER THE COURSE OF THE OCCULTATION. UNLIKE A TYPICAL ATMOSPHERIC OCCULTATION, THE JUNO POSITION WITHIN THE MAGNETOSPHERE CHANGES APPRECIABLY OVER THE COURSE OF A TORUS OCCULTATION. IS THIS MUCH OF AN ISSUE? WE NEED MARISSA TO GIVE A CRUDE REPRESENTATION OF MAGNETOSPHERIC DENSITY (I ALREADY HAVE SOMETHING THAT MIGHT WORK) AND WE ALSO NEED TO THINK A BIT ABOUT WHERE JUNO IS IN THE MAGNETOSPHERE DURING A TYPICAL OCCULTATION.
%DO YOU HAVE ANY ARTICLES ON EFFECTS OF INTERPLANETARY MEDIUM ON FREQUENCY? I FEEL WE'RE BEING TOO CAVALIER IN JUST BRUSHING IT OFF AS A BASELINE ISSUE. I FEEL SURE I'VE SEEN A FEW ARTICLES THAT DISCUSS THIS AND WHETHER IT IS A BIG ISSUE OR NOT.

The only remaining potentially significant source of error is Earth's ionosphere, which is discussed in Section~\ref{sec:earthiono}.

\subsection{\label{sec:earthiono}Initial model of Earth's ionosphere}

Plasma densities are much greater in Earth's ionosphere than in Jupiter's magnetosphere or the interplanetary medium.
Consequently, plasma in Earth's ionosphere can make a significant contribution to the line-of-sight column density despite the ionosphere's limited vertical extent. If plasma densities in Earth's ionosphere were constant along the line of sight over the duration of the occultation, then they would have no effect on the rate of change of the column density and would not affect the measured frequency shift.
This is commonly the case for radio occultation observations of planetary atmospheres and ionospheres, which last for minutes, not hours.
However, due to the large size of the IPT and the long duration of an IPT occultation, conditions in Earth's ionosphere along the line of sight from the ground station to the spacecraft may change appreciably over the course of the occultation.

The vertical column density, or vertical total electron content (TEC), of Earth's ionosphere varies with time of day, season, the solar cycle, and other factors \citep{2004Maruyama,2009Bagiya}. At nighttime, it can be represented by a constant value of 10 TECU from dusk until dawn.
After dawn, it increases smoothly to a peak value of $\sim$30 TECU at noon, then decreases smoothly to its nighttime value by dusk. 
This peak TEC of Earth's ionosphere, 30 TECU, is around 1.3 times the peak TEC of the IPT, 25.5 TECU.
%TEXT IS USING 1E16 UNITS WITHOUT EXPLAINING WHY A VALUE IS WRITTEN AS 30E16 NOT 3E17. NEED TO DEFINE TECU. OR USE NORMAL SCIENTIFIC NOTATION ON M-2 VALUES.
Moreover, line-of-sight TEC values will be greater than vertical TEC values by a factor of sec($\chi$), where $\chi$ is the zenith angle \citep{2004Mendilloa}.
%COMPARE TO GENERIC IPT VALUES FROM FIGURES.

\begin{figure}
\noindent\includegraphics[width=20pc]{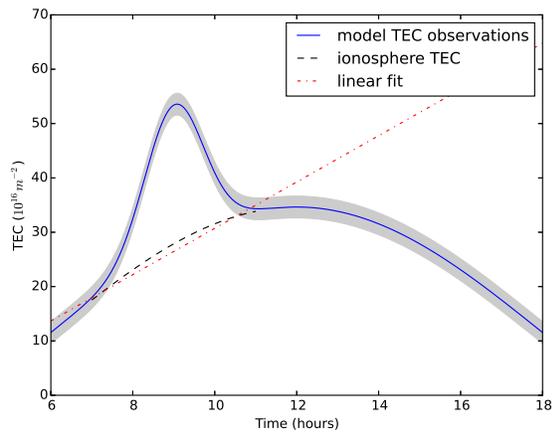}
\caption{Modeled line of sight total electron content between spacecraft and ground-based antenna during an occultation of the Io plasma torus as a function of local time at the ground-based antenna. The contribution of the Io plasma torus can be seen above the background contribution of Earth's ionosphere between 8 and 10 hours local time. The gray shaded region indicates the uncertainty on the measured total electron content. The black dashed line represents the modeled ionospheric TEC. The red dot-dashed line represents the linear fit to the background.}
%The solid curve in the figure shows the contributions of the ionosphere to the measurements of the total electron content over a period of 12 hours with the measurements starting at a time 6 hours. The gray shadded region shows the uncertainty on the measurements for total electron content. }
\label{fig:ionosphere}
\end{figure}

Figure~\ref{fig:ionosphere} illustrates how line-of-sight TEC through Earth's ionosphere and the IPT 
varies with time of day for a line-of-sight 30 degrees away from the zenith in which the radio signal passes through the center of the IPT at 9 hours local time.
The TEC is the sum of two components. The first component is from Earth's dayside ionosphere. It is given by 
$A + B \cos \left[ 2 \pi \left(LT - 12 \mathrm{hrs}\right) / \left(24 \mathrm{hrs}\right) \right]$, 
where $A$ equals 10 TECU, $B$ equals 20 TECU, and $LT$ is local time. 
The second component is from the IPT. It is given by 
$C \exp \left[-\left(LT - 9 \mathrm{hrs}\right)^{2} / \left(1 \mathrm{hr}\right)^{2} \right]$,
where $C$ equals 25.5 TECU and 1 hr equals 1 R$_J$/20 km s$^{-1}$ (Equation \ref{eqn:coldenssoln}). 
%The curve in Figure \ref{fig:ionosphere} is determined by adding the ionospheric TEC and the Io torus TEC together over the course of a day with the IPT being observed from 7--11 hours local time. 
We assume benchmark values of $N\left(0\right) = 2000$ cm$^{-3}$ and $H = 1 R_J$ for the IPT and a 36 second integration time.
Figure~\ref{fig:ionosphere} also shows representative uncertainties in TEC.
For conceptual simplicity, we neglect the variation in uncertainties with time that are defined by Equation~\ref{eqn:sqrtsigma} and adopt instead a constant uncertainty of 2 TECU. 
This value comes from the average of Equation \ref{eqn:uncertainty} over the assumed duration of the occultation. 
%XXX11 - Need to discuss that in reference to Eqn 11. 
The contributions of Earth's ionosphere to the measured line-of-sight TEC must be subtracted before properties of the IPT can be determined from the observations.
We consider two methods for doing so.

%Two methods can be considered with noise in the measurements corresponding to the uncertainty in a single measurement.

\begin{figure}
\noindent\includegraphics[width=20pc]{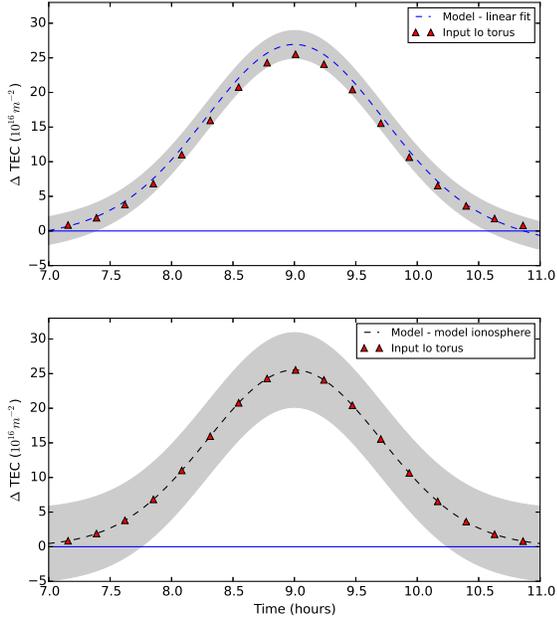}
\caption{(top) The dashed line shows the difference between the total electron content measurements from Figure \ref{fig:ionosphere} and a linear fit to background values either side of the occultation.
(bottom) The dashed line shows the difference between the total electron content measurements from Figure \ref{fig:ionosphere} and independent measurements of the total electron content in Earth's ionosphere. The differences shown in these panels are the inferred total electron contents of the Io plasma torus. The gray shaded regions show the uncertainties on the inferred total electron content of the Io plasma torus and the red triangles show the input total electron content of the Io plasma torus.}
\label{fig:noionosphere}
\end{figure}

First, we do a linear fit to the simulated measurements of TEC at 6--8 and 10--12 hours, then subtract this fit from the simulated measurements of TEC at 7--11 hours. 
The fit is shown as a red dot-dashed line in Figure \ref{fig:ionosphere}.
The residual TEC, which is shown in the top panel of Figure \ref{fig:noionosphere}, is the inferred contribution from the IPT.
This linear fitting method provides a baseline for the contributions of Earth's ionosphere. 
As can be seen in the top panel of Figure \ref{fig:noionosphere}, this method gives torus TEC values that are $\sim$ 1--2 TECU higher than the input torus TEC values from 7--11 hours. %WHAT DOES THE WORD PROFILE MEAN IN THIS CONTEXT? NOT DEFINED, NOT NECESSARY EITHER.
%A FEW MEANS THREE. DOESN'T LOOK THREE UNITS BIGGER. TEXT HAS NOT YET DEFINED TEC UNITS.
Although the corrected simulated measurements of torus TEC values are larger than the input torus TEC values, the difference is less than the measurement uncertainty of 2 TECU. 
Following Equation~\ref{eqn:coldenssoln}, we fit the corrected simulated measurements of torus TEC values to a function of the form of
$TEC\left(s\right) = TEC\left(0\right) e^{-\frac{s^2}{H^2}}$. %STATE FUNCTIONAL FORM.
The fitted peak TEC value is 27.21$\pm$0.06 TECU
and the fitted scale height, $H$, is 1.002$\pm$0.002 $R_J$. 
This peak TEC is 1.71 TECU larger than the input value of 25.5 TECU. Thus the fitted TEC value is 28 $\sigma$ away from the input TEC value, but the difference is only 7\% of the peak TEC.
The fitted scale height is 0.002 $R_J$ larger than the input value of $H$. 
The fitted scale height is 1 $\sigma$ away from the input scale height, but the difference is only 1\% of the scale height. %This difference is 1 $\sigma$, but $<$1\%.
The fitted peak TEC value and scale height imply a central density $N\left(0\right)$ of 2127.13 $\pm$ 6.33 cm$^{-3}$.
This inferred central density is 127.13 cm$^{-3}$ larger than the input value of 2000 cm$^{-3}$. 
The fitted central density value is 20 $\sigma$ away from the input central density value, but the difference is only 6\% of the density. %This difference is 20 $\sigma$, but only 6\%.
We conclude that this method is reasonable for subtracting the effects of Earth's ionosphere as the errors in the fitted IPT properties are less than 10\%. 
Yet it provides a poor characterization of the uncertainty in the fitted properties.
%LOOKS AS IF FITTED VALUE IS MANY SIGMA AWAY FROM TRUE VALUE, WHICH MUST BE DISCUSSED IN TEXT.%USE SANE NUMBER OF DECIMAL PLACES. COMPARE TO TRUE VALUE.  
%I NEED TO UNDERSTAND WHY TEXT HAS ASYMMETRIC UNCERTAINTIES. THIS IS SOMETHING THAT CAN BE FITTED WITH THE USUAL LEAST SQUARES PROCEDURES, WHICH GIVE SYMMETRIC ERRORS.
%SAME DISCUSSION FOR H.
%SAME COMPARISON AND DISCUSSION FOR N0.
%Performing a fit to the corrected profile we get a peak TEC and scale height of 27.21$^{+0.06}_{-0.05}$ TECU and 1.002$^{+0.002}_{-0.002}$ $R_J$.
%INTERPRETATION OF RESULTS. HOPEFULLY IPT EFFECTS ARE PRETTY OBVIOUS
%NEW FIGURE - RESIDUAL TEC VS TIME FROM 8 TO 10 HOURS. SUPERIMPOSED, TRUE IPT-ONLY TEC OVER SAME TIME INTERVAL.
%NEED TO THINK CAREFULLY ABOUT WHERE DIFFERENT UNCERTAINTIES ARE INTRODUCED AND CONSIDERED.
%SHOULD WE BE THINKING ABOUT THE ALLAN DEV UNCERT IN METHOD 1?
%PROBABLY ALSO WANT TO THINK ABOUT DOING A FIT TO RESIDUAL TEC, COMPARING FITTED N(0) AND H TO INPUT VALUES
%THAT WILL NEED UNCERTAINTIES, SURELY.

Second, we subtract modeled direct measurements of the contributions of Earth's ionosphere from the simulated measurements of line-of-sight TEC.
The ionospheric contribution is shown as a dashed black line in bottom of Figure \ref{fig:ionosphere}.
Here we assume Earth's ionospheric TEC follows the equation stated above for the dayside ionosphere, but that it is measured imperfectly.
GPS receivers at the NASA Deep Space Network (DSN) stations measure the TEC in Earth's ionosphere.
The TEC in Earth's ionosphere along the line of sight from the ground station to the spacecraft is routinely reported.
%in ASCII files. %TYPE OF DSN FILES AT HIGH TIME RESOLUTION. VERIFY IN THAT DSN EMAIL THAT STATED DIRECTION IS CORRECT.
%The typical uncertainty in these ionospheric TEC measurements is 2--5 TECU \citep{2000Thornton}.
%IS THIS UNCERTAINTY IN DATA OR STANDARD DEVIATION OF SOME MODEL? I DON'T SEE TEC DATA ON SWPC WEBSITE.
We subtract the contributions of Earth's ionosphere, which we assume to be known with an accuracy of 5 TECU \citep{2000Thornton}
%SO, ARE YOU USING THORNTON AS SOURCE FOR 5E16 OR THE SWPC WEBSITE? UNCLEAR ..., 
from the simulated measurements of TEC, which as before we assume to be known with an accuracy of 2 TECU.
The residual TEC, which is shown in the bottom panel of Figure \ref{fig:noionosphere}, is the inferred contribution from the IPT.
With this method, the corrected simulated measurements of TEC values match the input values well, whereas the values obtained with the first method were biased to larger values.
However, the measurement uncertainties are larger and thus the formal uncertainties on fitted parameters are also larger.
We fit the corrected simulated measurements of TEC values as above.
The fitted peak TEC value is 25.5$\pm$0.1 TECU, whereas the input value is 25.5 TECU. 
The difference between fitted and input peak TEC values is $<$1 $\sigma$.
The fitted scale height, $H$, is 1.005$\pm$0.007 $R_J$, whereas the input value is 1 R$_J$.
The difference between fitted and input scale heights is $<$1 $\sigma$.
The fitted peak TEC value and scale height imply a central density, $N\left(0\right)$, of 1994.18 $\pm$ 14.24 cm$^{-3}$, whereas the input central density is 2000 cm$^{-3}$. The difference between fitted and input central densities is $<$1 $\sigma$.
We conclude that this method is preferable. It accurately characterizes the fitted parameters. 
Furthermore the formal uncertainties are consistent with differences between fitted values and input values.

%SAME COMPARISON AND DISCUSSION FOR N0.
%The peak density and scale height of the Io plasma tours for this case are 25.5$^{+0.1}_{-0.1}$ TECU and 1.005$^{+0.007}_{-0.006}$ $R_J$.
%I NEED TO UNDERSTAND WHAT VARIABILITY MEANS IN THIS SENTENCE, IS IT JUST STANDARD DEVIATION OF SOMETHING OR WHAT?
%INTERPRETATION OF RESULTS. HOPEFULLY IPT EFFECTS ARE PRETTY OBVIOUS
%NEW FIGURE - RESIDUAL TEC VS TIME FROM 8 TO 10 HOURS. SUPERIMPOSED, TRUE IPT-ONLY TEC OVER SAME TIME INTERVAL.

%SUMMARY OF THIS SECTION
%I WILL WAIT UNTIL THE FITTED N0 VALUES ARE STATED AND I HAVE A BETTER UNDERSTANDING OF THE ERROR PROPAGATION BEFORE EDITING THIS PARAGRAPH.
%In this section we determine two distinct methods for determining the Io plasma torus profile in the presence of the Earth's daytime ionosphere. With the two methods we find that both methods result in an Io plasma torus profile that is relatively close to the true Io plasma torus parameters with exception of the scale height. From this we are able to determine the profile of the Io plasma torus even in the presence of the Earth's ionosphere. 

Having established the principle that the IPT can be observed using radio occultations despite the effects of Earth's ionosphere,
we neglect Earth's ionosphere in the remainder of this article.
More precisely, we assume that the observations occur during the nighttime such that the vertical TEC in Earth's ionosphere is relatively constant.
The contribution of Earth's ionosphere to the line-of-sight TEC can be found using either pre- or post-occultation observations, then subtracted from the TEC measurements.
%UNFORTUNATELY, THE LINE OF SIGHT CHANGES DIRECTION THROUGH THE NIGHT, SO WE CAN'T JUST SUBTRACT OFF A CONSTANT.
%NEED TO THINK ABOUT THIS ISSUE A BIT.

%where variations in the ionosphere are such that a constant can be subtracted for the Earth's total electron content. To determine the ability to calculate total electron content using radio occultations a model of the Io plasma torus using \textit{Voyager 1} data is created. From this model radio occultations of the Io plasma torus by the \textit{Juno} spacecraft are simulated.

\section{\label{sec:model}Sophisticated model of Io plasma torus}

Representing plasma densities in the IPT by a single Gaussian function is convenient and has been useful for testing the effects of changes in plasma and spacecraft parameters and effects of the Earth's ionosphere, but this representation oversimplifies the true density distribution in the IPT. %WHATEVER WE DID IN PREVIOUS SECTION

As discussed in Section \ref{sec:overview}, the IPT is conventionally divided into three regions: cold torus, ribbon, and warm torus. 
%The inner most region, with respect to Jupiter, is the cold torus. The cold torus is also characterized by being constrained to a thin region about the centrifugal equator (around 0.2 $R_J$ above and below the centrifugal equatorial plane).
%Moving out radially the next region we encounter is the ribbon. This region is characterized by high density of plasma in a region that is about 0.4 $R_J$ in radial extent and around 0.5 $R_J$ above and below the centrifugal equatorial plane. This region is located right before you reach the orbital distance of Io. 
%If we move still further out to the orbital distance of Io we reach the warm torus. The warm torus is the outermost region and extends from the orbit of Io all the way to the orbit of Europa. It is also the thickest region with an extent of around 2 $R_J$ above and below the centrifugal equatorial plane.
%NEED TO GIVE READER A SENSE OF WHAT THESE ARE LIKE WITHOUT JUST DUMPING VALUES OF PARAMETERS ON THEM. EG, INNER, OUTER; THIN, THICK.
%ALSO NEED TO THINK ABOUT ORDER IN LISTING THEM.
These three regions have distinct compositions, temperatures, and densities. 
To better understand temporal and spatial changes in the torus, it is desirable to measure densities in each of its constituent regions.
We therefore replace the single Gaussian function of Section \ref{sec:analytic} with a more sophisticated function that includes contributions for each region.

%we want to perform a fit that incorporates the differences in each of the regions. For this reason we fit the simulated profiles with three distinct functions rather than one function that covers the entire torus. 
%AT SOME POINT, POTENTIALLY RIGHT HERE, NEED TO EXPLAIN WHY WE WANT TO FIT IPT WITH THESE THREE THINGS NOT JUST ONE OVERALL GAUSSIAN

\subsection{Density distribution} %CHANGE SECTION NAME, CONSIDER WHAT'S A SECTION AND WHAT'S A SUBSECTION HERE}

We now represent the IPT by four functions, one each for the cold torus and ribbon, and two for the warm torus.
In the plane of the centrifugal equator, densities satisfy:

\begin{equation}
N(R < 6.1 R_{J}) = N_1 e^{- \frac{(R-C_1)^2}{(W_{1})^2}} + 
 N_2 e^{- \frac{(R-C_2)^2}{(W_{2})^2}} +  N_3 e^{- \frac{(R-C_3)^2}{(W_{3})^2}}
\label{eqn:equatorden}
\end{equation}
\begin{equation}
N(R > 6.1 R_J) = N_4 e^{- \frac{(R-C_4)^2}{(W_{4})^2}}
\label{eqn:fourth}
\end{equation}
where $R$ is distance away from the center of Jupiter in the equatorial plane. 
Equation~\ref{eqn:equatorden} contains three terms that represent the three regions of the torus: 1. cold torus, 2. ribbon, and 3. warm torus.
$N_1$, $N_2$, and $N_3$ correspond to the peak densities of the cold torus, ribbon, and warm torus components, respectively.
$C_1$, $C_2$, and $C_3$ are the central locations of the cold torus, ribbon, and warm torus components, respectively.
$W_1$, $W_2$, and $W_3$ are the radial widths, in $R_J$, of the cold torus, ribbon, and warm torus components, respectively.
Note that the total density at $R = C_1$, say, is the sum of the three terms. It is not simply $N_1$.

The warm torus is not well-represented by a single term, which is why Equation~\ref{eqn:equatorden} only applies at $R < 6.1 R_J$.
At larger radial distances, the plasma density is given by Equation~\ref{eqn:fourth}.
We label this region as the extended torus. It has peak density $N_4$, central location $C_4$, and radial width $W_4$.

In order to extend this model beyond the plane of the centrifugal equator, we multiply each term in Equations~\ref{eqn:equatorden}--\ref{eqn:fourth} by  factor of $e^{-\frac{r^2}{H^2}}$ where $r$ is distance away from the plane of the centrifugal equator.
Therefore $N\left(R, r\right)$ satisfies:

\begin{eqnarray}
N(R< 6.1 R_J, r) = N_1 e^{- \frac{(R-C_1)^2}{(W_{1})^2}}e^{- \frac{r^2}{H_1^2}} + 
 N_2 e^{- \frac{(R-C_2)^2}{(W_{2})^2}}e^{- \frac{r^2}{H_2^2}} +  N_3 e^{- \frac{(R-C_3)^2}{(W_{3})^2}}e^{- \frac{r^2}{H_3^2}} 
 \label{eqn:density1}\\
N(R > 6.1 R_J, r) = N_4 e^{- \frac{(R-C_4)^2}{(W_{4})^2}}e^{- \frac{r^2}{H_{3}^2}}
\label{eqn:density2}
\end{eqnarray}

$H_1$, $H_2$, $H_3$, and $H_{3}$ are the scale heights of the cold torus, ribbon, warm torus, and extended torus components, respectively.
Note that the warm torus and extended torus have the same scale height, $H_{3}$.

\begin{figure}
\noindent\includegraphics[width=20pc]{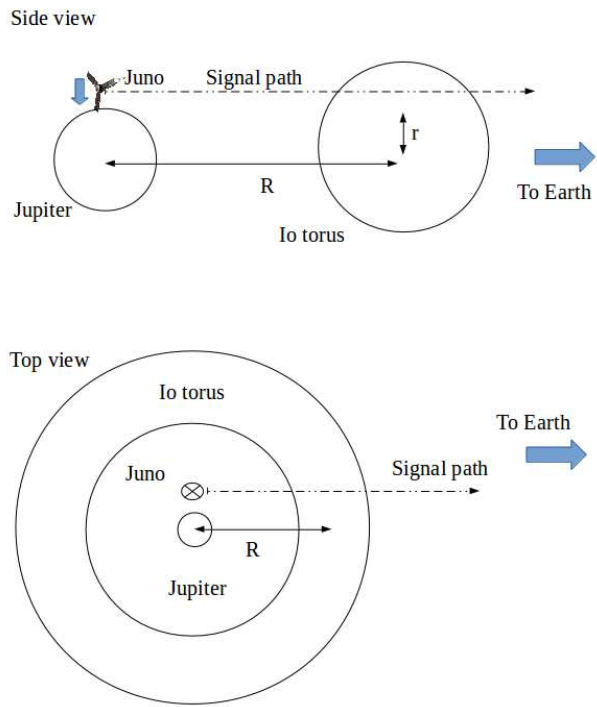}
\caption{Schematic of the sophisticated model of the Io plasma torus and the geometry of an occultation.
The electron density in the Io plasma torus is a function of the position coordinates $r$ and $R$.
(top) View from the dawn side of Jupiter with the Earth to the right. The arrow beside \textit{Juno} shows the spacecraft's direction of motion. (bottom) View looking down on the north pole of Jupiter with Earth to the right. Here the \textit{Juno} spacecraft is moving into the page.}
\label{fig:geomschem}
\end{figure}

\begin{figure}
\noindent\includegraphics[width=20pc]{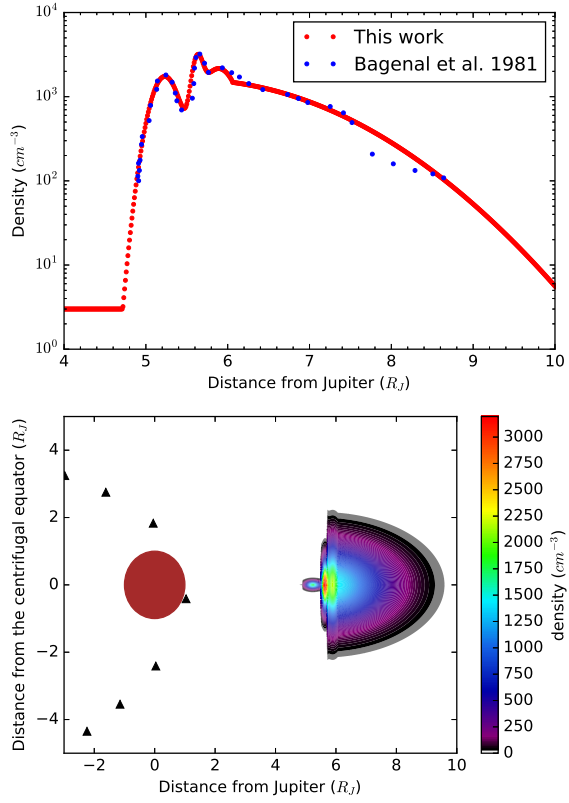}
\caption{(top) Red symbols show electron densities in the plane of the centrifugal equator as a function of distance from Jupiter for the sophisticated model of the Io plasma torus. The innermost peak corresponds to the cold torus, the intermediate peak to the ribbon, and the outermost peak to the warm torus. The abrupt change in gradient at 6.1 $R_{J}$ corresponds to the transition from the warm torus to the more distant extended torus.
The blue points show \textit{Voyager 1} data from \citet{1981Bagenal}.
(bottom) Colors indicate electron densities in the sophisticated model of the Io plasma torus. The red disk indicates Jupiter and the black triangles mark the position at one hour intervals of \textit{Juno}, simulated by the University of Iowa ephemeris tool (www-pw.physics.uiowa.edu/$\sim$jbg/juno.html).}
%This figure shows the model of the Io Plasma torus created from the \textit{Voyager} data published in \citet{1981Bagenal}. The top panel shows the plasma distribution as a function of distance from Jupiter seen in the centrifugal equator. The bottom panel is the plasma distribution as a function of distance from Jupiter and distance above and below the centrifugal equator(cold torus scale height:0.1 $R_{J}$, ribbon scale height: 0.25 $R_{J}$, and warm torus scale height: 1.0 $R_{J}$). The triangles mark the position at one hour intervals of \textit{Juno}, simulated by the University of Iowa ephemeris tool (www-pw.physics.uiowa.edu/~jbg/juno.html). The circle in the middle marks the location of Jupiter ($1R_J$).}
\label{fig:model}
\end{figure}

The functional form represented by Equations~\ref{eqn:equatorden}--\ref{eqn:fourth} was adopted in order to reproduce the radial density distribution for the centrifugal equator shown in Figure 6 of \citet{1981Bagenal}. Numerical values of the corresponding model parameters, which were determined by a fit to the data shown in that figure, are given in Table~\ref{tab:modelparam}. 
Numerical values of the model scale heights, which were determined from Figure 12 in \citet{1981Bagenal}, are given in Table~\ref{tab:modelparam}. 
A schematic of the model IPT and the occultation geometry is shown in Figure \ref{fig:geomschem}.
The modeled electron densities are shown in Figure \ref{fig:model}.
Figure \ref{fig:model} also demonstrates that this model provides a good representation of the density observations in the centrifugal equator reported in 
Figure 6 of \citet{1981Bagenal}.
%\textbf{REPEATING THE SAME FIGURE TWICE WITH MINOR DIFFERENCES IS NOT DESIRABLE. INCLUDE THE COMPARISON TO BAGENAL IN THE TOP PANEL OF THE ORIGINAL FIGURE.}

%Equation~\ref{eqn:scaleheight}, with ion temperatures from \citet{2004Thomas}, and ion compositions discussed in Section~\ref{sec:overview}.
%CAN'T SHOW THIS FIGURE UNTIL YOU'VE COMPLETED THE DESCRIPTION OF THE MODEL.
%I HAVE ZERO CLUE HOW DISTANCE R ALONG IO-JUPITER LINE IS USED IN THESE FUNCTIONS. TABLE MENTIONS A THING CALLED WIDTH, BUT THERE'S NOTHING IN THE TEXT ABOUT THAT DIMENSION.
%DEFINE MODEL HERE.
%TEXT MENTIONS A FOURTH GAUSSIAN BRIEFLY, BUT WITHOUT EXPLAINING HOW IT IS USED (IN MODEL GENERATION OR IN FIT TO SIMULATED DATA?) OR WHAT ITS PARAMETERS ARE.

\subsection{Simulated \textit{Juno} radio occultation}

\begin{figure}
\noindent\includegraphics[width=20pc]{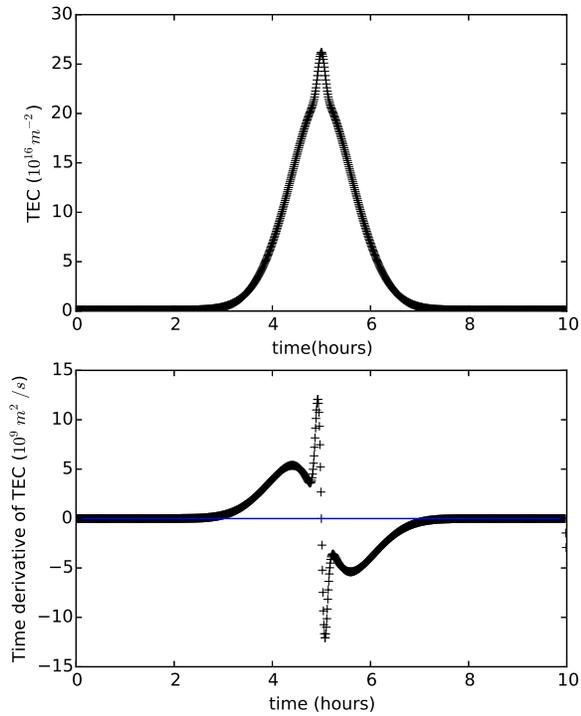}
\caption{(top) Total electron content from the model shown in Figure \ref{fig:model} as function of time. At the start time, the spacecraft is approximately 4 $R_J$ above the plane of the centrifugal equator. The subtle changes in the slope of the curve between 4 and 6 hours are signatures of the cold torus and ribbon.
%WHICH SUBTLE CHANGES? NEED TO EITHER HIGHLIGHT ON PLOT WITH ARROWS OR STATE TIMES IN CAPTION.
(bottom) Numerically calculated rate of change of the total electron content shown in the top panel.}
%Shown here is the total electron content (TEC)[top] and the time derivative [bottom] for the simulation. This is for the ideal case without noise added. The horizontal line in the bottom plot is the line denoting zero.}
\label{fig:idealtec}
\end{figure}

\begin{figure}
\noindent\includegraphics[width=20pc]{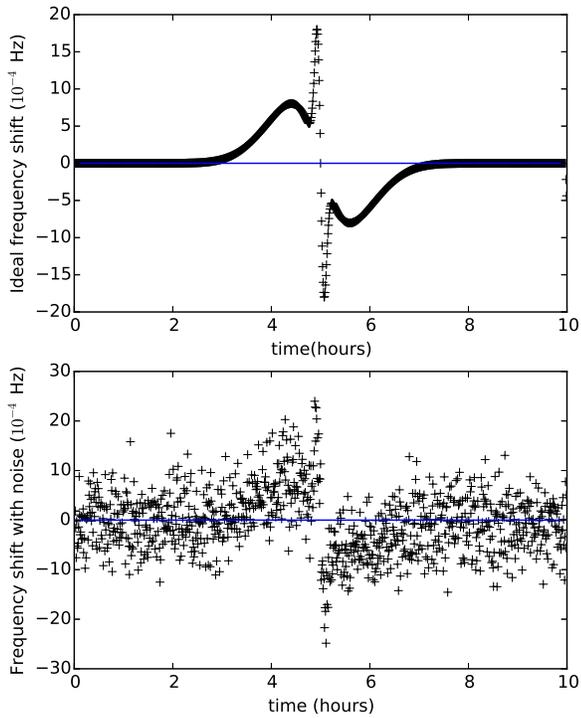}
\caption{(top) Noise-free frequency shift calculated using Equation \ref{eqn:shiftfreq} and the rate of change of total electron content shown in the bottom panel of Figure \ref{fig:idealtec} as function of time.
(bottom) Same as top panel, but with noise added to the received frequencies.}
%Shown here is the shift in frequency calculated from the ideal total electron content [top] and the frequency with noise at the $3x10^{-14}$f level [bottom], where f is the frequency used. The shift is shown in units of $10^{-4} Hz$. }
\label{fig:shiftfrequency}
\end{figure}

To simulate a radio occultation through this representation of the IPT, we assume that the line of sight from \textit{Juno} to Earth is parallel to the centrifugal equator. %THIS IS NECESSARY FOR US TO BE ABLE TO REDUCE THE SIMULATION TO MOTION NORMAL TO THIS PLANE. ALSO FOR TEC(T) TO INVOLVE INTEGRAL OVER BIG-R AT FIXED LITTLE-R.
We assume that the spacecraft velocity in the direction normal to the centrifugal equatorial plane is -20 km s$^{-1}$,
assume that nodding motion of the IPT due to Jupiter's rapid rotation can be neglected,
and use an integration time of 36 seconds, which corresponds to a sampling rate of 0.03 Hz.
%LANGUAGE ISSUE - SPEED IS A POSITIVE DEFINITE QUANTITY, SO CAN'T BE NEGATIVE. SAME BASIC PROBLEM AS +/-S IN EARLIER SECTION.
%I PRESUME SAME SAMPLING RATE IS USED THROUGHOUT, WHICH MEANS ALL MY 10 SECOND STUFF EARLIER NEEDS TO BE FIXED.
Figure \ref{fig:idealtec} shows the TEC and its rate of change. 
Figure \ref{fig:shiftfrequency} shows the corresponding noise-free frequency shift $\Delta f$ (Equation \ref{eqn:shiftfreq}) and the noisy frequency shift $\Delta f$. 
Following Section \ref{sec:analytic}, for relative measurement uncertainties of $3 \times 10^{-14}$ on $f_{R,X}$ and $f_{R,Ka}$, the uncertainty on a single measurement of $\Delta f$ is $3.8 \times 10^{-4}$ Hz. 
The uncertainties are added to the frequency shifts 
pulling from a random normal distribution with mean zero and standard deviation of $3.8 \times 10^{-4}$ Hz.
%SUDDENLY 3.828 HZ HAS CHANGED TO 4 HZ WITHOUT ANY DISCUSSION.
%UNCERTAINTY VALUE NEEDS TO BE SCALED UP VIA USUAL ERROR PROPAGATION FORMULA.

\begin{figure}
\noindent\includegraphics[width=20pc]{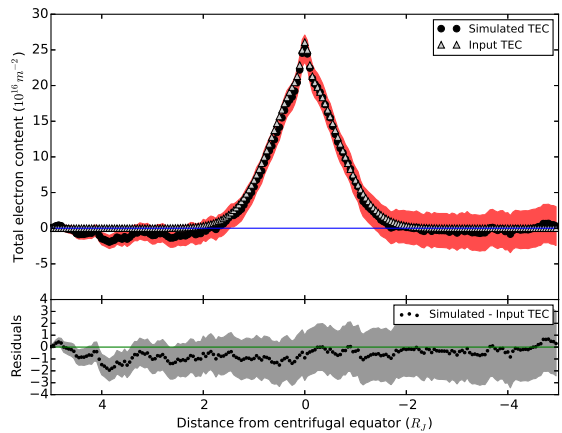}
\caption{(top) The black points show the total electron content found by integrating the frequency shift in the bottom panel of Figure \ref{fig:shiftfrequency}. The red shaded region is the uncertainty on the values. The gray triangles show the input total electron content values from the top panel of Figure \ref{fig:idealtec}. The blue line marks the location of zero.
(bottom) The black symbols show the residuals between the simulated and the input total electron content. The gray shaded region marks the uncertainty on the residuals. The green line marks the location of zero.}
%Shown here is the total electron content with noise(black points) with the uncertainty level(lighter outline) determined by propagation of error through the equations using standard procedures. The noise is $3x10^{-14}$ times the frequency.}
\label{fig:tecnoise}
\end{figure}

The simulated measurements of TEC were found by integration of the frequency shift $\Delta f$ using Equation \ref{eqn:shiftfreq}.
Uncertainties in the TEC were derived from the uncertainty in $\Delta f$ 
by repeated application of the standard error propagation formula.
The top panel of Figure \ref{fig:tecnoise} shows the simulated measurements of TEC, corresponding uncertainties, and the input TEC.
The bottom panel shows the difference between simulated measurements of TEC and the input TEC.
%MODIFY FIGURE 9 TO INCLUDE A RESIDUAL PANEL BELOW, LIKE FIGURE 10.
%RESIDUAL PANEL SHOWS MEASURED MINUS ACTUAL WITH GREY SHADING INDICATING UNCERTAINTY ON MEASURED.

It is noticeable that the relative uncertainties on the TEC (Figure \ref{fig:tecnoise}) are much less than those on the frequency shift (Figure \ref{fig:shiftfrequency}) from which TEC was derived. This is an example of integration reducing the importance of random noise.

\subsection{Fitted Io plasma torus parameters and their accuracy}

Section \ref{sec:analytic} explored the accuracy with which a central density and scale height could be fit to simulated TEC observations.
However, this used a simple single Gaussian model of the IPT.
Here we fit the simulated TEC measurements from Section \ref{sec:model} to a model that includes multiple Gaussian contributions in order to determine the accuracy with which the central density and scale height of the cold torus, ribbon, and warm torus can be measured.

For clarity in this initial exploration of this topic, we assume that the radio occultation is observed at nighttime.
During the night, the TEC of Earth's ionosphere is relatively constant.
%WELL, EXCEPT FOR THAT DRATTED CHANGING ZENITH ANGLE SEC CHI EFFECT. THIS TEXT CAN BE DELETED ONCE WE FIGURE OUT HOW TO PHRASE THE RELEVANT DISCUSSION AT END OF PREVIOUS SECTION.
A constant TEC will have no effect on the measured frequency shift (Equation \ref{eqn:shiftfreq}).
Consequently we neglect the effects of Earth's ionosphere and fit the simulated TEC observations shown in Figure \ref{fig:tecnoise}.

Since the line-of-sight between the spacecraft and Earth is assumed to be parallel to the plane of the centrifugal equator, each radio ray path has a constant value of $r$.
The model TEC along the ray path with closest approach distance $r$ is derived in Appendix~\ref{app:tec}. It satisfies:

\begin{eqnarray}
\label{eqn:erf04again}
TEC\left(r\right) =
\sqrt{\pi} N_{1} W_{1} e^{- \frac{r^2}{H_{1}^2}} +\\ \nonumber
\sqrt{\pi} N_{2} W_{2} e^{- \frac{r^2}{H_{2}^2}} + \\ \nonumber
\frac{\sqrt{\pi}}{2} \left[
N_{3} W_{3} 
\left( 1 + \frac{ \left( 6.1 R_{J} - C_{3} \right) }{W_{3}} \right) +
N_{4} W_{4} 
\left( 1 - \frac{ \left( 6.1 R_{J} - C_{4} \right) }{W_{4}} \right) 
\right]
e^{- \frac{r^2}{H_{3}^2}} 
\end{eqnarray}

Due to their different scale heights, the three regions of the IPT each make distinct and potentially separable contributions to the overall TEC.
We therefore fit the simulated TEC observations to a function of the form:
\begin{equation}
TEC(r) = A e^{- \frac{r^2}{B^2}} + 
 Ce^{-\frac{r^2}{D^2}} +  Ee^{-\frac{r^2}{F^2}}
\label{eqn:fitequation}
\end{equation}
The parameters $A$, $C$, and $E$ corresponds to the peak or equatorial TEC for each of the regions and the parameters $B$, $D$, and $F$ correspond to the scale heights of the cold torus ($H_{1}$), ribbon ($H_{2}$), and warm torus ($H_{3}$), respectively.
%IN THIS EQUATION, IS N ELECTRON DENSITY OR TEC? IF ELECTRON DENSITY, NEED TO EXPLAIN HOW TEC IS FIT TO AN EQUATION THAT DOES NOT CONTAIN TEC.
%IF TEC, THAT EXPLAINS WHY THIS EQUATION DEPENDS ON Z BUT NOT RADIAL DISTANCE R. IF ELECTRON DENSITY, NEED TO EXPLAIN WHY R VANISHES.
%IF TEC, NEED TO EXPLAIN WHY GAUSSIANS OF Z ARE APPROPRIATE TO FIT THE N(R,Z) FUNCTIONS THAT DEFINE THE MODEL.
%NEED TO EXPLAIN PHYSICAL CONNECTIONS BETWEEN PARAMETERS A TO F AND WHATEVER PARAMETERS DEFINE N(R,Z) IN THE MODEL.

\begin{figure}
\noindent\includegraphics[width=20pc]{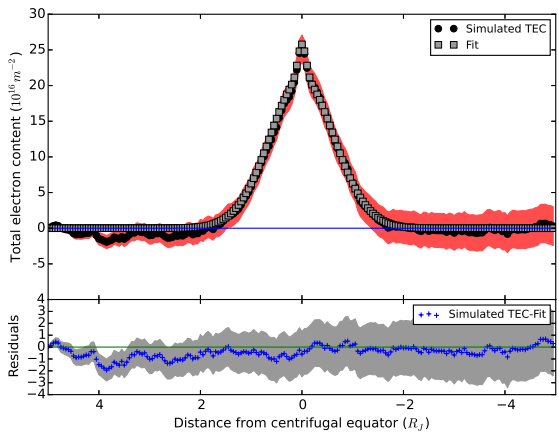}
\caption{(top) The black points show the total electron content found by integrating the frequency shift in the bottom panel of Figure \ref{fig:shiftfrequency}. The red shaded region is the uncertainty on the values. The gray squares show the MCMC fit to the integrated total electron content. The blue line marks the location of zero.
(bottom) The blue plus signs show the residual between the fit and the simulated total electron content. The gray shaded region marks the uncertainty on the residuals. The green line marks the location of zero.}
%This figure shows the MCMC fit and its residuals. The top plot shows the fit (gray squares), the input total electron content (gray triangles), and the simulated data with error (black circles with red outline). The bottom plot shows the residuals between data and fit (blue +) and ideal data and fit (black circles). The blue line in the top plot and the green line in the bottom plot gives the zero line in both plots. }
\label{fig:tecfit}
\end{figure}

We fit this equation to the simulated TEC observations shown in Figure \ref{fig:tecfit} using a Markov Chain Monte Carlo (MCMC) method.
This is implemented using the Python module emcee, which is an open source MCMC ensemble sampler developed by \citet{2013Foreman}.
%I DO NOT UNDERSTAND THIS PARAGRAPH.
%To get the starting parameters minimize the negative likely-hood for each parameter in equation \ref{eqn:fitequation}. This uses the probability function, which is assumed to be Gaussian for each parameter, and takes the negative natural logarithm and finds where thethe true value. The parameters with the highest probability of being the true values are used a minimal value. This finds the value of the parameter with the highest probability of being s the initial guess for the MCMC code. The code runs through a number of fits and outputs . 

Figure~\ref{fig:tecfit} shows the simulated measurements and fitted TEC, as well as the residuals between the simulated measurements and the fit.
Table \ref{tab:fitparam} shows the best fit parameters for each region compared to the input values. 
Two of the three fitted peak electron content values are within 1 $\sigma$ of their input values, and the other is within 2 $\sigma$, which demonstrates that they are reliable. All three fitted scale heights are within 10\% and 1 $\sigma$ of their input values.
%The fitted peak electron content values are within 10\% of their input values, which demonstrates that such fitted values would have scientific value.
%The fitted peak electron content values are within 2 $\sigma$ of the input values, which demonstrates that they are reliable.
%COLD TORUS PEAK TEC IS NOT WITHIN 1 SIGMA
%The fitted scale heights are also within 10\% and 2 $\sigma$ of their input values.

%Each true value of the parameters are within the one sigma level for each of the fitted parameters except for the scale height of the ribbon. For the ribbon scale height the true value is more than three sigma from the fitted value which could be due to an underestimation of the errors. Thus, the parameters of the Io plasma torus can be derived with reasonable uncertainty in the values. 

%The MCMC method produces a best fit parameter distribution that can be plotted for each parameter along with the covariance of the parameters (Figure \ref{fig:paramspace}). 
%NEED TO DRAW SOME IMPLICATIONS FROM THIS FIGURE, OTHERWISE THERE IS NO POINT TO SHOWING THE FIGURE.
%NEED TO MENTION TABLE 3 AND DISCUSS ITS IMPLICATIONS.
%NEED TO ADD TRUE VALUES OF PARAMETERS TO TABLE 3.
%NEED TO GET N0 FROM TEC AND H FOR EACH REGION AS WELL (FITTED WITH ERRORS AND TRUE VALUE) $DONE IN DISCUSSION$

%ALL THE FIGURE CAPTIONS NEED EDITING. I AM LEAVING THAT UNTIL THE TEXT IS SETTLED.
%BTW, THE PARAMETER VALUES IN FIGURES 1 AND 2 APPEAR TO HAVE BEEN SELECTED AT RANDOM. USE ROUND NUMBERS INSTEAD.

\section{\label{sec:discussion}Discussion}

The preceding sections showed how radio signals from the \textit{Juno} spacecraft could be used to measure TEC profiles for the IPT, that uncertainties on measured TEC are relatively small, and that a fit to the measured TEC can determine the scale height and peak TEC for each of the three regions of IPT (cold torus, ribbon, and warm torus).

Ion temperatures can be derived from scale heights via Equation \ref{eqn:scaleheight}.
We assume that S$^{+}$ dominate in the cold torus, O$^{+}$ dominates in the ribbon, and S$^{2+}$ and O$^{+}$ dominates in the warm torus such that the mean molecular mass is 24 daltons \citep{2004Thomas}. % Assuming these compositions are consistent for all observations then the scale height can be thought of as equivalent to the ion temperature. 
We use the best fit parameters and uncertainties reported in Table~\ref{tab:fitparam} to find 
ion temperatures of 0.957$^{+0.173}_{-0.173}$ eV for the cold torus, 16.7$^{+1.58}_{-2.47}$ eV for the ribbon and 56.9$^{+6.05}_{-5.51}$ eV for the warm torus. 
For reference, the ion temperatures reported by \citet{2004Thomas} and discussed in Section \ref{sec:overview} are
1--4 eV for the cold torus, 10--30 eV for the ribbon, and $\approx$60 eV for the warm torus.
Hence the fitted ion temperatures are reasonable for the cold torus, ribbon, and the warm torus.

%for all regions except the ribbon. This we contribute to a potential miss interpretation of the value of the scale height for the ribbon from the literature. For example, if we say that the ribbon scale height is 0.62 $R_J$ (taken from density curves in \citet{1997Bagenal}), rather than the 0.25 $R_J$ we have specified, then we arrive at a more reasonable temperature of 30.4 eV for the ions.  
%XXX13 - We need to clear this point up.
%This method gives values close to those quoted in \citet{2004Thomas} and mentioned in section \ref{sec:overview} for all regions except the ribbon. This we contribute to a potential miss interpretation of the value of the scale height for the ribbon from the literature. For example, if we say that the ribbon scale height is 0.62 $R_J$, rather than the 0.25 $R_J$ we have specified, then we arrive at a more reasonable temperature of 9.8 eV for the ions.  
%XXX14 - What happened to the factor of sqrt(pi) from Eqn 15? Text here shuffles between peak and average confusedly.

The peak or equatorial TEC of each region can be determined by fitting Equation~\ref{eqn:fitequation} to the observed $TEC\left(r\right)$.
For the cold torus and ribbon, peak TEC equals $\sqrt{\pi} N_{i} W_{i}$, where $N_{i}$ is the maximum density in region $i$ and $W_{i}$ is the width of region $i$.
For the warm torus and its extension beyond 6.1 $R_{J}$, peak TEC is more complicated (Equation~\ref{eqn:erf04again}). Nevertheless, it can be considered as the product of a maximum density and an effective width.
If the width of a region is known from independent measurements or models of the IPT, then the maximum density for that region can be found from the observed peak TEC.
As noted by \citet{1992Bird}, the electron density in a region cannot be accurately determined from an observed peak TEC without independent knowledge of the width and central peak location of that region.
The analysis described in this article assumes that the line of sight from \textit{Juno} to Earth is parallel to the plane of the centrifugal equator.
If that is not the case, then the measured TEC values would correspond to cuts through the torus at the angle between the line of sight and the centrifugal equator. This is equivalent to the IPT being tilted.
A tilted torus can be accounted for by a suitable adjustment of the assumed Gaussian profile, as in the model by \citet{1983Divine}. 

\section{\label{sec:conclusion}Conclusions}
When the line of sight between \textit{Juno} and Earth passes through the Io plasma torus, which occurs once per orbit, 
radio signals from the \textit{Juno} spacecraft can be used to measure total electron content profiles for the Io plasma torus.
We develop a model of densities in the Io plasma torus using values measured by the \textit{Voyager 1} spacecraft and reported in
\citet{1981Bagenal}, then use it to simulate a dual-frequency 
radio occultation performed using the telecommunication subsystem on the \textit{Juno} spacecraft. 
Using the modeled densities we calculate the total electron content by integrating along a line of sight parallel to the torus equator. 
From the total electron content we are able to derive the frequency shift that would be measured by the Deep Space Network receiving stations. 
This is then used with error introduced equal to the Allan deviation corresponding to an integration time on the order of 10s to determine a simulated profile of the measured total electron content. 

Uncertainties on the measured total electron content are relatively small ($\sim$10\%).
A Markov chain Monte Carlo fit to the measured total electron content can determine the scale height and peak total electron content for each of the three regions of Io plasma torus (cold torus, ribbon, and warm torus). The ion temperature in each region can be determined from the scale height assuming independent knowledge of the ion composition.
The peak total electron content in each region is proportional to the product of the peak local electron density and the region's width in the equatorial plane. However, without independent knowledge of one of these two factors, the other cannot be determined directly.
Numerical modeling of the Io plasma torus may be useful in narrowing the range of possible peak local electron densities and widths.
%It has been shown that the scale height and peak plasma density can be derived from a fit to simulations of radio occultations performed using the telecommunication subsystem on the \textit{Juno} spacecraft. The radio occultation uses the Ka-band and X-band radio frequencies during the spacecraft's closest approach to Jupiter. During closest approach the plasma of the Io plasma torus will pass between the receiver and the spacecraft and the signal passing through this plasma will be refracted. This refraction leads to the frequencies measured on the ground being shifted from the transmitted frequencies. From these shifted frequencies we derive total electron content for the Io plasma torus. These profiles are fit with a sum of Gaussian functions to extract the scale height and peak total electron content values for the three regions of the torus. From the scale height we can derive the temperature of the ions using equation \ref{eqn:scaleheight} and the peak total electron content is used to calculate the peak electron density using $N0*width$. Thus the shifted frequencies of the \textit{Juno} spacecraft can inform us about properties of the Io plasma torus. 
%SKETCH OUT THE SCIENCE SCENARIO WE DISCUSSED OF GROUND-BASED IR VOLCANIC ACTIVITY OBSERVATIONS, GROUND-BASED SODIUM CLOUD OBSERVATIONS, JUNO IPT RADIO OCCULTATIONS, ALL THE OTHER JUNO OBSERVATIONS PROVIDING AN INTEGRATED, COMPREHENSIVE VIEW OF THE FLOW OF STUFF FROM IO'S VOLCANOES TO JUPITER AURORAL EMISSIONS. 

To date, only two radio occultations of the Io plasma torus have been performed, \textit{Voyager 1} \citep{1979Eshlemanb} and \textit{Ulysses} \citep{1992Bird}.
\textit{Juno} has the potential to perform over 20 occultations. This series of occultations would provide a rich picture of the structure of the Io plasma torus and its temporal and spatial variability.

The \textit{Juno} mission presents an unparalleled opportunity to study the flow of material from the volcanoes of Io to the auroral regions of Jupiter with simultaneous observations of all stages in this system.
%During the tenure of the \textit{Juno} mission there will arise the opportunity for the study of the flow of material from the volcanism and atmosphere of Io to the auroral emissions. 
Ground-based infrared observations of Io can be used to monitor the moon's volcanic activity \citep{2014deKleer}.
Ground-based sodium cloud observations can be used to monitor the transport of material from Io's atmosphere into the neutral clouds, since sodium can be considered as a tracer for sulfur and oxygen \citep{2002Wilson,2004Mendillob,2004Thomas}. 
Radio occultations can be used to monitor the ionization of neutral species and the distribution of plasma within the Io plasma torus \citep{1979Eshlemanb,1992Bird}.
\textit{Juno}'s suite of plasma instruments will monitor plasma densities in the acceleration regions near Jupiter's poles \citep{2014Bagenal}.
Together, the measurements already planned by the \textit{Juno} mission, the potential radio occultations of the Io plasma torus, and Earth-based observations of the Jupiter system will reveal the complete life-cycle of plasma in Jupiter's magnetosphere.

%XXX16 - Make all JGR references consistent format

%%% End of body of article:

\appendix

\section{\label{app:tec}Total electron content}

Since the line-of-sight between the spacecraft and Earth is assumed to be parallel to the plane of the centrifugal equator, each radio ray path has a constant value of $r$.
The total electron content along the ray path with closest approach distance $r$, $TEC\left(r\right)$, satisfies:

\begin{eqnarray}
\label{eqn:erf01}
TEC\left(r\right) =
N_{1} W_{1} \frac{\sqrt{\pi}}{2}
\left( erf \left[ \frac{C_{1}}{W_{1}} \right]
+ erf \left[ \frac{ \left( 6.1 R_{J} - C_{1} \right) }{W_{1}} \right] \right) 
e^{- \frac{r^2}{H_{1}^2}} + \\ \nonumber \\ \nonumber
N_{2} W_{2} \frac{\sqrt{\pi}}{2}
\left( erf \left[ \frac{C_{2}}{W_{2}} \right]
+ erf \left[ \frac{ \left( 6.1 R_{J} - C_{2} \right) }{W_{2}} \right] \right) 
e^{- \frac{r^2}{H_{2}^2}} + \\ \nonumber \\ \nonumber
N_{3} W_{3} \frac{\sqrt{\pi}}{2}
\left( erf \left[ \frac{C_{3}}{W_{3}} \right]
+ erf \left[ \frac{ \left( 6.1 R_{J} - C_{3} \right) }{W_{3}} \right] \right) 
e^{- \frac{r^2}{H_{3}^2}} + \\ \nonumber \\ \nonumber
N_{4} W_{4} \frac{\sqrt{\pi}}{2}
\left( erf \left[ \frac{C_{4}}{W_{4}} \right]
- erf \left[ \frac{ \left( 6.1 R_{J} - C_{4} \right) }{W_{4}} \right] \right) 
e^{- \frac{r^2}{H_{3}^2}} 
\end{eqnarray}

where $erf\left(x\right)$ is the error function.
For all plausible conditions, $C_{1}/W_{1}$, $C_{2}/W_{2}$, $C_{3}/W_{3}$, and $C_{4}/W_{4}$ are much greater than one.
Since $erf\left(x \gg 1\right) = 1$, Equation~\ref{eqn:erf01} becomes:

\begin{eqnarray}
\label{eqn:erf02}
TEC\left(r\right) =
N_{1} W_{1} \frac{\sqrt{\pi}}{2}
\left( 1
+ erf \left[ \frac{ \left( 6.1 R_{J} - C_{1} \right) }{W_{1}} \right] \right) 
e^{- \frac{r^2}{H_{1}^2}} + \\ \nonumber \\ \nonumber
N_{2} W_{2} \frac{\sqrt{\pi}}{2}
\left( 1
+ erf \left[ \frac{ \left( 6.1 R_{J} - C_{2} \right) }{W_{2}} \right] \right) 
e^{- \frac{r^2}{H_{2}^2}} + \\ \nonumber \\ \nonumber
N_{3} W_{3} \frac{\sqrt{\pi}}{2}
\left( 1
+ erf \left[ \frac{ \left( 6.1 R_{J} - C_{3} \right) }{W_{3}} \right] \right) 
e^{- \frac{r^2}{H_{3}^2}} + \\ \nonumber \\  \nonumber
N_{4} W_{4} \frac{\sqrt{\pi}}{2}
\left( 1
- erf \left[ \frac{ \left( 6.1 R_{J} - C_{4} \right) }{W_{4}} \right] \right) 
e^{- \frac{r^2}{H_{3}^2}} 
\end{eqnarray}

Furthermore, $\left(6.1 R_{J} - C_{1}\right)/W_{1}$ and $\left(6.1 R_{J} - C_{2}\right)/W_{2}$ can also be expected to be greater than one, which gives:

\begin{eqnarray}
\label{eqn:erf03}
TEC\left(r\right) =
\sqrt{\pi} N_{1} W_{1} e^{- \frac{r^2}{H_{1}^2}} + \\ \nonumber \\ \nonumber
\sqrt{\pi} N_{2} W_{2} e^{- \frac{r^2}{H_{2}^2}} + \\ \nonumber \\ \nonumber
N_{3} W_{3} \frac{\sqrt{\pi}}{2}
\left( 1
+ erf \left[ \frac{ \left( 6.1 R_{J} - C_{3} \right) }{W_{3}} \right] \right) 
e^{- \frac{r^2}{H_{3}^2}} + \\ \\ \nonumber
N_{4} W_{4} \frac{\sqrt{\pi}}{2}
\left( 1
- erf \left[ \frac{ \left( 6.1 R_{J} - C_{4} \right) }{W_{4}} \right] \right) 
e^{- \frac{r^2}{H_{3}^2}} 
\end{eqnarray}

In our model, $\left(6.1 R_{J} - C_{3}\right)/W_{3} = 0.66$ and 
$\left(6.1 R_{J} - C_{4}\right)/W_{4} = 0.30$.
The error function $erf\left(x\right)$ increases from 0 at $x=0$ to 1 at $x \gg 1$.
It can be approximated as $erf\left(x\right) = x$ for $x < 1$ and 
$erf\left(x\right) = 1$ for $x > 1$.
The error in this approximation is less than 0.15 for all $x$.
We therefore assume that $\left(6.1 R_{J} - C_{3}\right)/W_{3} < 1$ and
$\left(6.1 R_{J} - C_{4}\right)/W_{4} < 1$, which leads to:

\begin{eqnarray}
\label{eqn:erf04}
TEC\left(r\right) =
\sqrt{\pi} N_{1} W_{1} e^{- \frac{r^2}{H_{1}^2}} + 
\sqrt{\pi} N_{2} W_{2} e^{- \frac{r^2}{H_{2}^2}} + \\ \nonumber \\ \nonumber
\frac{\sqrt{\pi}}{2} \left[
N_{3} W_{3}
\left( 1 + \frac{ \left( 6.1 R_{J} - C_{3} \right) }{W_{3}} \right) +
N_{4} W_{4}
\left( 1 - \frac{ \left( 6.1 R_{J} - C_{4} \right) }{W_{4}} \right) 
\right]
e^{- \frac{r^2}{H_{3}^2}} 
\end{eqnarray}

Expanding the term in square brackets further does not provide additional insight.

%%%%%%%%%%%%%%%%%%%%%%%%%%%%%%%%
%% Optional Appendix goes here
%
% \appendix resets counters and redefines section heads
% but doesn't print anything.
% After typing \appendix
%
%\section{Here Is Appendix Title}
% will show
% Appendix A: Here Is Appendix Title
%
%%%%%%%%%%%%%%%%%%%%%%%%%%%%%%%%%%%%%%%%%%%%%%%%%%%%%%%%%%%%%%%%
%
% Optional Glossary or Notation section, goes here
%
%%%%%%%%%%%%%%
% Glossary is only allowed in Reviews of Geophysics
% \section*{Glossary}
% \paragraph{Term}
% Term Definition here
%
%%%%%%%%%%%%%%
% Notation -- End each entry with a period.
% \begin{notation}
% Term & definition.\\
% Second term & second definition.\\
% \end{notation}
%%%%%%%%%%%%%%%%%%%%%%%%%%%%%%%%%%%%%%%%%%%%%%%%%%%%%%%%%%%%%%%%
%
%  ACKNOWLEDGMENTS

\begin{acknowledgments}
PHP was supported, in part, by the Massachusetts Space Grant Consortium (MASGC). 
PHP would also like to thank Mark Veyette and Paul Dalba for useful discussions.
We would like to thank the two anonymous reviewers for their suggestions. 
No data were used in this article. 
%about the code and fit used for this project.  
\end{acknowledgments}

\end{article}
%
%
%% Enter Figures and Tables here:
%
% DO NOT USE \psfrag or \subfigure commands.
%
% Figure captions go below the figure.
% Table titles go above tables; all other caption information
%  should be placed in footnotes below the table.
%
%----------------
% EXAMPLE FIGURE
%
% \begin{figure}
% \noindent\includegraphics[width=20pc]{samplefigure.eps}
% \caption{Caption text here}
% \label{figure_label}
% \end{figure}
%\clearpage

\clearpage

\clearpage

%\begin{figure}
%\noindent\includegraphics[width=20pc]{Htest.eps}
%\caption{Dependence of frequency shift on closest approach distance $s$ for initial model of torus. Different lines represent different values for $H$.}
%Shown here is how the shift in frequency is affected by different plasma scale heights. Since r is a function of time the x-axis can be thought of as time increasing to the right.}
%\label{fig:Htest}
%\end{figure}

%\clearpage

%\begin{figure}
%\noindent\includegraphics[width=20pc]{Vtest.eps}
%\caption{Dependence of frequency shift on closest approach distance $s$ for initial model of torus. Different lines represent different values for $ds/dt$.
%XXX18 - Change drdt in legend to dsdt.
%}
%Shown here is how the shift in frequency is affected by different plasma scale heights. Since r is a function of time the x-axis can be thought of as time increasing to the right.}
%\label{fig:Vtest}
%\end{figure}

%\clearpage

%\begin{figure}
%\noindent\includegraphics[width=20pc]{test_fit.eps}
%\caption{This figure shows the test of the fit to the data given by the single Gaussian representation of the Io plasma torus.}
%\label{fig:testfit}
%\end{figure}

%\clearpage

\clearpage

\clearpage

\clearpage

\clearpage

\clearpage

\clearpage

\clearpage

\clearpage

%\begin{figure}
%\noindent\includegraphics[width=35pc]{papererrorfig.eps}
%\caption{This figure shows the triangle plot of the MCMC fit to the simulated total electron content. The diagonal plot shows the shape of the probability distributions for the fitted parameters. The lines crossing the plots correspond to the true input values of the simulation.}
%\label{fig:paramspace}
%\end{figure}

%\clearpage

%\begin{figure}
%\noindent\includegraphics[width=35pc]{region_division.eps}
%\caption{This figure shows where each region adds to the full total electron content profile.}
%\label{fig:regionpieces}
%\end{figure}
%
%\clearpage
%
%
% ---------------
% EXAMPLE TABLE
%
%\begin{table}
%\caption{Time of the Transition Between Phase 1 and Phase 2\tablenotemark{a}}
%\centering
%\begin{tabular}{l c}
%\hline
% Run  & Time (min)  \\
%\hline
%  $l1$  & 260   \\
%  $l2$  & 300   \\
%  $l3$  & 340   \\
%  $h1$  & 270   \\
%  $h2$  & 250   \\
%  $h3$  & 380   \\
%  $r1$  & 370   \\
%  $r2$  & 390   \\
%\hline
%\end{tabular}
%\tablenotetext{a}{Footnote text here.}
%\end{table}

\begin{table}
\caption{\textit{Juno} radio signal parameters \citep{2012Mukai}}
\centering
\begin{tabular}{l c c}
\hline
    \multicolumn{2}{c}{Parameter}   &  Value    \\
\hline\\
  Downlink Frequency [GHz] & \\
  & X-Band  & 8.40413    \\
  \\
  & Ka-Band  & 32.0833 \\
  Turnaround Ratio & \\
  & X to X-Band & $\frac{880}{749}$ \\
  \\
  & X to Ka-Band & $\frac{3344}{749}$\tablenotemark{a} \\
  \\
  & $\frac{f_{D,X}}{f_{D,Ka}}$  & $\frac{880}{3344}$ \\
  \\
\hline
\end{tabular}
\label{tab:Junoparamtable}
\tablenotetext{a}{Assumed to match Cassini radio system \citep{2004kliore}.}
\end{table}

\clearpage

\begin{table}
\caption{Input parameters for Io plasma torus model.}
\centering
\begin{tabular}{l c c c c}
\hline
   Region & Central Density  & Scale Height & Peak Location & Width \\
   & [cm$^{-3}$] & [$R_J$] & [$R_J$] & [$R_J$] \\ 
\hline
   \\ 
   Cold Torus  & 1710 & 0.1 & 5.23 & 0.20  \\
   \\
   Ribbon  & 2180 & 0.6 & 5.60  & 0.08 \\
   \\
   Warm Torus  & 2160 & 1.0 & 5.89  & 0.32 \\
   \\
   Extended Torus & 1601  & 1.0 & 5.53 & 1.88 \\ 
   \\
\hline
\end{tabular}
\label{tab:modelparam}
\tablenotetext{}{The central density, peak location, and width of the regions are derived from Figure 6 of \citet{1981Bagenal}. The scale heights are derived from Figure 12 of \citet{1981Bagenal}. Observations from the \textit{Galileo} and \textit{Cassini} spacecraft suggest that the ribbon position changes over time, but in general the center is thought to be located between 5.5 and 5.9 $R_J$ \citep{2004Thomas}.}
\end{table}

\clearpage

\begin{table}
\caption{Best fit parameters from the MCMC fit to the simulated TEC data.}
\centering
\begin{tabular}{l c c c c}
\hline
   Parameter & Region  & Fit Value & Input Value  \\
\hline
\\
  Central TEC [$10^{16}$ m$^{-2}$] & Cold Torus  & $4.67^{+0.32}_{-0.20}$ & 4.29  \\ \\
  & Ribbon  & $3.98^{+0.25}_{-0.21}$ & 3.79 \\ \\
  & Warm Torus  & $17.06^{+0.79}_{-0.83}$ & 17.17\\ \\
  Scale Height [$R_J$] & Cold Torus  & $0.11^{+0.01}_{-0.01}$ & 0.10  \\ \\
  & Ribbon  & $0.65^{+0.03}_{-0.05}$ & 0.60 \\ \\
  & Warm Torus  & $0.98^{+0.06}_{-0.05}$ & 1.0 \\ \\
  Reduced Chi-Squared &  1.004  \\
\hline
\end{tabular}
\label{tab:fitparam}
%\tablenotetext{a}{Footnote text here.}
\end{table}

% See below for how to make sideways figures or tables.

\end{document}